\documentclass[showpacs,amssymb,eqsecnum,aps]{revtex4}
\usepackage{amsmath}
\usepackage{graphicx}
\usepackage{epsf}
\usepackage{graphics}
\usepackage[latin1]{inputenc}
\newcommand{\be}{\begin{equation}}
\newcommand{\ee}{\end{equation}}
\newcommand{\bea}{\begin{eqnarray}}
\newcommand{\eea}{\end{eqnarray}}

\newcommand{\nnbb}{\nonumber\\}

\newcommand{\tris}{\sum\!\!\!\!\!\!\!\!\!\!\int \!\!\!\!\!\!\int \!\!\!\!\!\!\int} 
\newcommand{\lar}{\leftrightarrow}

\newcommand{\dst}{\displaystyle}

\begin{document}

\title{On the Free Energy of Noncommutative Quantum Electrodynamics at High Temperature}

\author{F. T. Brandt, J. Frenkel and  C. Muramoto}
\affiliation{Instituto de Física, Universidade de São Paulo,
  05508-090, São Paulo, SP, BRAZIL}

\begin{abstract}
We compute higher order contributions to the free energy
of noncommutative quantum electrodynamics at a 
nonzero temperature $T$. Our calculation includes up to 
three-loop contributions (fourth order in the coupling constant $e$).
In the high temperature limit we sum all the {\it ring diagrams} and
obtain a result which has a peculiar dependence on the 
coupling constant. For large values of $e\theta T^2$ 
($\theta$ is the magnitude of the noncommutative parameters)
this non-perturbative contribution exhibits a non-analytic behavior 
proportional to $e^3$. We show that above a certain
critical temperature, there  occurs a thermodynamic
instability which may indicate a phase transition. 
\end{abstract}


\maketitle

\section{Introduction}
This paper is about the thermodynamics of quantum electrodynamics
formulated in a noncommutative space (NCQED), which constitutes
a system of self-interacting gauge fields 
\cite{Szabo:2001kg,Hayakawa:1999yt}.
We employ the imaginary time formalism \cite{kapusta:book89} 
in order to compute the {\it free energy} of the pure 
gauge sector of the theory. Previous investigations on
this subject  have revealed interesting properties already at the 
lowest non trivial order (two-loop order) \cite{Arcioni:1999hw}.
The main purpose of the present paper is to take into account the higher order
corrections to the free energy. 


Corrections to the free energy, higher than two-loops, in thermal field theories 
forces us to take into consideration the sum of an infinite series of
diagrams. This happens because the fields acquire, from interactions,
an effective thermal mass. This so-called {\it plasmon effect} is
known to happen in gauge theories, like SU(N), even without fermions
(pure Yang-Mills theory) because of the non-Abelian character 
of fields \cite{Kapusta:1979fh}. Inasmuch NCQED is a gauge theory 
of self-interacting gauge fields, it
is natural to consider the possibility of similar non-perturbative effects
when we take into account higher order corrections to the free energy.
As we will see the non-perturbative effect is more subtle 
in the case of NCQED because the would be infrared singularities are smoothed by the
noncommutative scale (this would not be the case in the noncommutative
version of the U(N) theory). However, in the present work we will show
that in the regime of temperatures
much higher than the inverse of the noncommutative scale, we meet a
breakdown of the perturbative series and the need for summing an
infinite series of diagrams is ineluctable.


In order to pinpoint the breakdown of the perturbative regime we first 
analyze in detail the two-loop contributions. We perform the
calculation in a general co-variant gauge and obtain a gauge
independent result in terms of a relatively simple function of the 
temperature which can be computed analytically
for both the low and high temperature regimes. We also plot the result
for all intermediate values of the temperature.
We show explicitly that the two-loop result converges to a finite limit
when the temperature $T$ is much higher than the inverse of the
noncommutative scale $1/\sqrt{\theta}$, or, equivalently 
when $\tau\equiv \theta T^2\gg 1$. 


Then we consider all the three-loop contributions and we verify that
similarly to what happens in the commutative fields theories, the
so-called {\it ring diagram} is dominant in the high temperature regime,
being proportional to $\tau$. We also show that this set of graphs, 
as well as the higher order ring diagrams are independent
of the gauge fixing parameter. We find that the relative strength of
the ring diagrams is of order $(e\tau)^2$, where $e$ is the coupling constant.
After summing all the rings, the resulting expression is
non-analytic in the coupling and behaves as $e^3$ for 
large values of $e\tau$. We also investigate the behavior of the free
energy as a function of $e\tau$ and show that, 
above some critical temperature $T_c$, there occurs a thermodynamic 
instability. This is manifested through the appearance of an imaginary
part in the free energy, which can be directly related to the
decay rate of a metastable vacuum \cite{Affleck:1980ac}.
This behavior is induced by the presence of a noncommutative magnetic
mode in the theory, which is associated with the transverse component
of the static self-energy.
We show that the magnetic mass associated
with the noncommutative transverse mode, for temperatures close to the critical temperature,
is proportional to $(T_c^2-T^2)^{1/2}$.


In the next section we present the basic features of NCQED as well as
our notation and conventions. In Sec. III we compute the two-loop
contributions to the free energy and consider the limits of high and
low temperature up to sub-leading terms. In Sec IV we compute the three-loop contributions 
and obtain the dominant high temperature behavior in terms of graphs
involving self-energy  insertions. In Sec. V we
compute the sum of all ring diagrams, investigate the properties of
the free energy and obtain the critical value of the temperature.
In the last section we discuss the main results and the connection
between the instability and the noncommutative magnetic mode.
We leave to the appendices the technical details of the rather
involved calculations of the Feynman graphs.

\section{noncommutative QED}
The gauge invariance of the QED action 
\be\label{ncqedaction}
S = -\frac{1}{4} \int{\rm d}^d x F_{\mu\nu}(x) \star F^{\mu\nu}(x).
\ee
under a $U$($1$) gauge transformation implies that the gauge potentials,
in terms of which the electromagnetic field $F_{\mu\nu}$ is defined,
must have self-interactions when the theory is formulated on a
noncommutative manifold \cite{Hayakawa:1999yt}.
The basic reason for this is that the usual product of two functions, $f(x)\, g(x)$
is replaced by the Grönewold-Moyal star product \cite{Szabo:2001kg}.
\begin{eqnarray}\label{moyal1}
f(x)\star g(x) = f(x)~\exp\left(\frac i2\,\overleftarrow{\partial_\mu}\,
\theta^{\mu\nu}\,\overrightarrow{\partial_\nu}\right)~g(x) ,
\end{eqnarray}
where $\theta^{\mu\nu}=-\theta^{\nu\mu}$ has canonical dimension of
inverse square mass and satisfies
\be
[x^\mu,x^\nu] = i \theta^{\mu\nu}.
\ee
Since the $U$($1$) transformation acts on the gauge fields in such a
way that
\be
A^\prime_\mu = \frac{i}{e} U\star \partial_\mu U^{-1} + U\star A_\mu \star U^{-1}  ,
\ee
the extra terms in the variation of the action can be compensated using
\be
F_{\mu\nu} = \partial_\mu A_\nu - \partial_\nu A_\mu 
- i e \left[A_\mu, A_\nu \right]_{\star},
\ee
where $\left[f(x), g(x) \right]_{\star} \equiv f(x)\star g(x) - g(x)\star f(x)$.
Consequently, in the noncommutative version of QED (NCQED) there are cubic and quartic
interaction vertices similarly to the case of the Yang-Mills theory. Of
course the details of the interaction vertices are quite different
from the usual gauge theories since there are no color charges in NCQED.

The quantization of this theory follows closely the usual approach
employed in the case of non-Abelian gauge fields. In an co-variant
class of gauges one adds the gauge fixing and ghost actions given
respectively by
\be\label{gf}
S_{\rm gf} = -\frac{1}{2\xi}\int{\rm d}^d x
\left(\partial_\mu A^\mu\right)\star\left(\partial_\nu A^\nu\right)
\ee
and
\be\label{ghost}
S_{\rm ghost} = \int{\rm d}^d x\,
\partial^\mu \bar c \star \left(\partial_\mu c 
-i e [A_\mu,c]_\star\right).
\ee
The full quantum behavior can be described in terms 
the sum of (\ref{ncqedaction}),  (\ref{gf})  and (\ref{ghost}) which yields
the following  Feynman rules for the propagators and the interaction vertices 
(for convenience we are not including the the factor $i$ 
from $\exp{(i\,S)}$ in the path integral)  
\begin{subequations}
\begin{eqnarray}\label{1f1gen}
\includegraphics[scale=0.85]{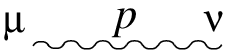}
& : \;\;\;&\;\;\; \displaystyle{
{1\,\over p^{2}+i\epsilon} \left(\eta_{\mu\nu} -
 (1-\xi){p_{\mu}p_{\nu}\over p^{2}}\right)} \label{1f1} \\
\includegraphics[scale=0.7]{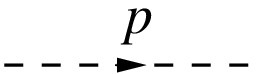}
 & : \;\;\;&\;\;\; 
\displaystyle{-{1\,\over p^{2} + i\epsilon}}\label{1fgen2}
\end{eqnarray}
\begin{eqnarray}\label{1f2}
\begin{array}{c}\includegraphics[scale=0.8]{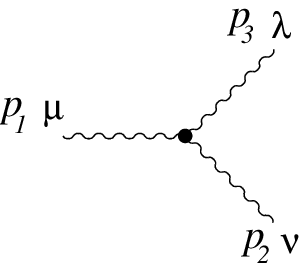}\end{array}
 & :\;\;\;  &\;\;\;  -2\,i\,e\,\sin\left(\frac{p_1\times p_2}{2}\right)
\left[(p_{1}-p_{2})^{\lambda}\eta^{\mu\nu} +
      (p_{2}-p_{3})^{\mu}\eta^{\nu\lambda} +
      (p_{3}-p_{1})^{\nu}\eta^{\lambda\mu}\right] \nnbb  & & \\ 
\begin{array}{c}\includegraphics[scale=0.8]{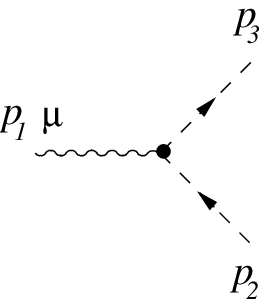}\end{array}
 & :\;\;\;  &\;\;\;  -2\,i\,e\,\sin\left(\frac{p_2\times p_3}{2}\right)\, p_{3}^{\mu}
 \nnbb & & \\ 
\begin{array}{c}\includegraphics[scale=0.8]{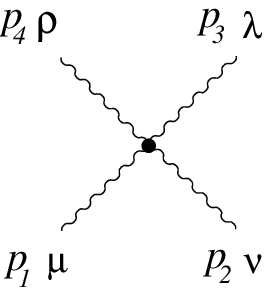}\end{array}
 & :\;\;\;  &\;\;\;  -4\, e^{2}\left[
\sin\left(\frac{p_1\times p_2}{2}\right)\,\sin\left(\frac{p_3\times p_4}{2}\right)
(\eta^{\mu\lambda}\eta^{\nu\rho} - \eta^{\mu\rho}\eta^{\nu\lambda}) 
\right.\nonumber\\
 &  & \qquad\quad + 
\sin\left(\frac{p_1\times p_3}{2}\right)\,\sin\left(\frac{p_4\times p_2}{2}\right)
(\eta^{\mu\rho}\eta^{\lambda\nu}-\eta^{\mu\nu}\eta^{\lambda\rho})
\nonumber\\ & & \nonumber\\ & & \nonumber\\
 &  & \qquad\quad\left. + 
\sin\left(\frac{p_1\times p_4}{2}\right)\,\sin\left(\frac{p_2\times p_3}{2}\right)
(\eta^{\mu\nu}\eta^{\rho\lambda} - \eta^{\mu\lambda}\eta^{\rho\nu}) 
\right],\nnbb & & \label{1vertex}
\end{eqnarray}
\end{subequations}
({\it all momenta} are inward and Dirac delta functions for the
conservation of momenta are understood). Our Minkowski metric convention is  $(+\,-\,-\,\cdots\,-)$
so that $p^2=p_0^2-|\vec p|^2$. The wavy and dashed lines
represent respectively the gauge and ghost fields and we have introduced
the notation 
\be
{p\times q} \equiv p_\mu\,\theta^{\mu\nu}\,q_\nu.
\ee

In order to study the thermodynamics of NCQED we will employ the
imaginary time formalism. This amounts to replace the zero
components of all momenta by $2\pi\, i \, n\, T$ ($n=0,\pm 1,\pm 2\,\cdots$).
Then the $d$-dimensional integration over the loop momenta is modified according to the rule
\be\label{TermVac1}
\int{\rm d}^d p f(p_0,\vec p) \rightarrow
T\displaystyle{\int {\rm d}^{d-1} p \sum_{n=-\infty}^{\infty}}
f(p_{0}=i\,\omega_n,\vec p),
\ee
where $\omega_n \equiv 2\pi\, n\, T$.

Another important consistency requirement is that $\theta^{i0}=0$. 
Then the momentum dependence in the interaction vertices depends only
on the spacial components ($p\times q= p_i \theta^{ij} q_j$) so that
the sum over the Matsubara frequencies $\omega_n$ can be performed
using the standard techniques. As we will see in the following
sections the non-commutativity of space coordinates 
has interesting consequences on the thermodynamical properties of the system.


\section{The lowest order contributions to the free energy}
Most of the results of this section can also be found in reference \cite{Arcioni:1999hw}.
One reason to reproduce these results here is because we
need to specify our notation and conventions as well as our 
basic approach. We will show how our approach allows to obtain 
in a simple and straightforward way the two-loop result in terms of a relatively
simple function of the temperature which can be explicitly computed for both
the low and high temperature regimes. In particular we obtain an
explicit result for the sub leading contributions at high $\tau$.

Let us start by defining clearly what we are going to compute.
Our first main goal is to obtain the free energy per unit of volume 
\cite{kapusta:book89} 
\be
\tilde \Omega(T,\theta) =\frac{\Omega(T,\theta)}{V} = -\frac{T}{V} \log Z(T, \theta)
\ee
of NCQED theory defined in the previous section. Here $Z(T,\theta)$ is
the partition function, which in the imaginary time formalism, has the form
\be
Z(T, \theta) = \int {\cal D} A \,{\cal D}\bar c\,{\cal D} c
\, \exp{\left[-\left(S+S_{\rm gf}
+S_{\rm ghost}\right)\right]}
\ee
where the actions inside the exponential are given by the Eqs. 
(\ref{ncqedaction}),  (\ref{gf})  and (\ref{ghost}) with the replacement
$x_0\rightarrow -i x_4$ and $x_4$ integrated from zero to $1/T$.
The argument $\theta$ represents the dependence of 
the partition function on the noncommutative parameter $\theta_{\mu\nu}$.

\begin{figure}[h!]
\includegraphics[scale=0.8]{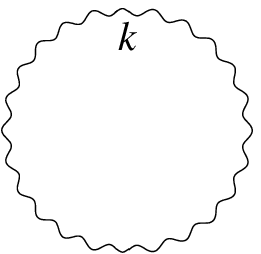}\qquad\includegraphics[scale=0.8]{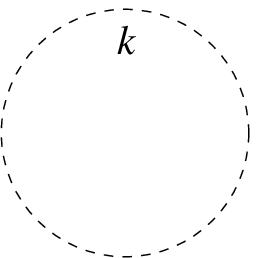}
\caption{One-loop diagrams which contribute to the free-energy.
Wavy and dashed lines denote respectively vector and ghost fields.} \label{fig0}
\end{figure}

The lowest order contribution to $\Omega(T,\theta)$ can be represented
diagrammatically by the graphs shown in the figure \ref{fig0}. This is
the free-field contribution which of course is independent of $\theta$ 
(the noncommutative character only shows up trough the interaction 
vertices in Eqs. (\ref{1f2}) ). 
This gives the known result of free QED, which, in $d=4$ yields
\be\label{free0}
\frac{\Omega^{(0)}(T)}{V} = 2\,\int \frac{{\rm d}^3 k}{(2\pi)^3} 
\left[\frac{|\vec k|}{2} + T\,\log\left(1-{\rm e}^{-(k/T)}\right)\right].
\ee
Notice that the factor $2$ on the right side of the above equation
counts the number of physical degrees of freedom of the massless gauge
boson $A_\mu$ (the ghost loop contribution in figure \eqref{fig0}
cancel the unphysical degrees of freedom).
The first term inside the square bracket of Eq. (\ref{free0})
represents the temperature independent (infinite) contribution of the
zero point energy of the vacuum. Subtracting this contribution, 
and performing an integration by parts, we obtain
\be\label{free01a}
\tilde\Omega^{(0)}(T) =
\frac{\Omega^{(0)}(T)}{V} = -\frac{1}{3\,\pi^2}\,\int_0^\infty 
{k^3 N_B(k)\;  {\rm d} k}  , 
\ee
where 
\be\label{1defbose1}
N_B(k) \equiv \frac{1}{{\rm e}^{(k/T)}-1} 
\ee
is the Bose-Einstein thermal distribution. Using the formula \cite{gradshteyn}
\be\label{1boseint1}
\int_0^{\infty}\frac{x^{n-1}}{{\rm e}^{x} - 1} {\rm d} x = \Gamma(n) \zeta(n)
\ee
where $\zeta$ is the Riemann zeta function, yields 
\be\label{free01}
\tilde\Omega^{(0)}(T) = -\frac{\pi^2}{45} \, T^4.
\ee

\begin{figure}[h!]
\includegraphics[scale=0.8]{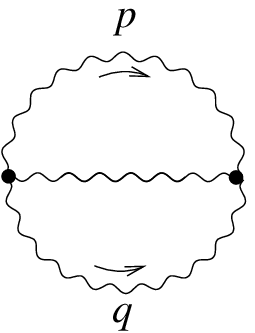}\qquad\includegraphics[scale=0.8]{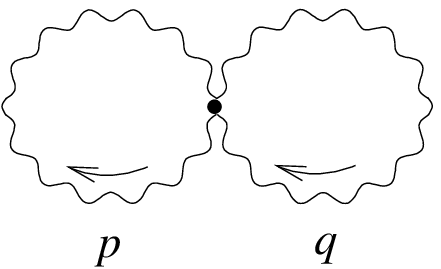}\qquad\includegraphics[scale=0.8]{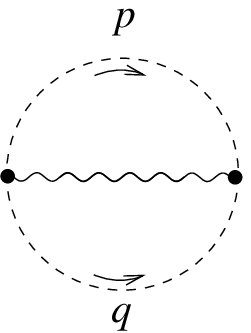}
\caption{Two-loop diagrams which contribute to the free-energy
in NCQED.} \label{1fig2}
\end{figure}

The first effects of interactions appears at ${\cal O}(e^2)$ and can be
computed from the diagrams in figure \ref{1fig2}. 
The individual contributions of these diagrams (including the symmetry
factors and the minus sign associated with the ghost loop) are
\begin{equation}\label{1tl1}
\frac{1}{12}
\begin{array}{c}\includegraphics[scale=0.8]{I1}\end{array}=
e^2\,\sum\!\!\!\!\!\!\!\!\!\int{\rm d} p
\,\sum\!\!\!\!\!\!\!\!\!\int{\rm d} q\, I_1 ,
\end{equation}
\begin{equation}
\frac{1}{8}
\begin{array}{c}\includegraphics[scale=0.8]{I2}\end{array}=
e^2\,\sum\!\!\!\!\!\!\!\!\!
\int{\rm d} p\,\sum\!\!\!\!\!\!\!\!\!
\int{\rm d} q\, I_2
\end{equation}
and
\begin{equation}
-\frac{1}{2}
\begin{array}{c}\includegraphics[scale=0.8]{I3}\end{array}=
e^2\,\sum\!\!\!\!\!\!\!\!\!
\int{\rm d} p\,\sum\!\!\!\!\!\!\!\!\!\int
{\rm d} q\, I_3 .
\end{equation}
We have introduced the compact notation
\be
\sum\!\!\!\!\!\!\!\!\!\int {\rm d} p= 
T\displaystyle{\int \frac{{\rm d}^{d-1} p}{(2\pi)^{d-1}} 
\sum_{n=-\infty}^{\infty}}.
\ee
Using the Feynman rules given in Eqs. (\ref{1f1}) and (\ref{1f2})
the integrands $I_1$, $I_2$ and $I_3$ can be expressed as
\begin{equation}
I_1 = -  (d-1)\left(\frac{1}{p^2\,q^2}+
                            \frac{1}{p^2\,(p+q)^2}+
                            \frac{1}{q^2\,(p+q)^2}\right)
\,\sin^2\left(\frac{p\times q}{2}\right) ,
\end{equation}
\begin{equation}
I_2 =   (d-1)\,d\,\frac{1}{p^2\,q^2}\,\sin^2\left(\frac{p\times q}{2}\right) ,
\end{equation}
and
\begin{equation}
I_3 =  \left(\frac{1}{p^2\,q^2}+
                    \frac{1}{p^2\,(p+q)^2}-
                    \frac{1}{q^2\,(p+q)^2}\right)
\,\sin^2\left(\frac{p\times q}{2}\right).
\end{equation}
Since $I_1$, $I_2$ and $I_3$ are integrands of dimensionally
regularized integrals, one can perform shifts in order to simplify
the above expressions before the computation of the integrals and
sums. In this way we can simplify $I_1$ and $I_3$ performing the shifts
$q\rightarrow q - p$ and $p\rightarrow p - q$ in their second and third
terms respectively.  Noticing that the $\sin^2$-factor 
remains unchanged under these shifts, all the terms are
reduced to a single type of
momentum dependence which has the same form of $I_2$.
Combining the reduced integrands we obtain
\begin{equation}\label{1fe2}
\frac{1}{12}
\begin{array}{c}\includegraphics[scale=0.6]{I1}\end{array}+
\frac{1}{8}
\begin{array}{c}\includegraphics[scale=0.6]{I2}\end{array}-
\frac{1}{2}
\begin{array}{c}\includegraphics[scale=0.6]{I3}\end{array}=
e^2\,\left(d-2\right)^2\, T^2\,
\sum\!\!\!\!\!\!\!\!\!\int{\rm d} p\, \sum\!\!\!\!\!\!\!\!\!\int{\rm d} q\,
\sin^2\left(\frac{p\times q}{2}\right)\frac{1}{p^2 q^2} 
\end{equation}
The factor $(d-2)^2$ is to be expected from the gauge invariance of $\Omega(T,\theta)$.
Indeed, if we employ an axial gauge condition in two dimensions so
that $A^0 = \kappa\,A^1$ ($\kappa$ is some constant) the theory becomes 
free because $[A^0,A^1]=0$. In general the gauge invariance of higher
order contributions can be rather subtle because of the gauge dependence of
the running coupling constant $e$. However, we have computed
Eq. (\ref{1fe2}) using the vector field propagator in Eq. (\ref{1f1})
showing explicitly that the result does not 
depend on $\xi$, as we show in the Appendix (\ref{appB}).
This can be understood taking into account that the relation between the
coupling constant computed in two different gauges 
$e^2(\xi=1)=e^2(\xi) + {\cal O}(e^4(\xi))$ does not change the $e^2$
contribution in Eq. (\ref{1fe2}). This however is not the case of
contributions of higher order. Even so, the 
factor $d-2$ can be present in the higher order contributions 
because the coupling constant would not be modified by quantum corrections
in the free two-dimensional theory. 

The computation of the sums and integrals in Eq. (\ref{1fe2}) can be
performed in a straightforward way. Using the standard relation
\cite{kapusta:book89} 
\be\label{1TermVac}
T\displaystyle{\sum_{n=-\infty}^{\infty}}f(p_{0}=i\,\omega_n)=
{             {\int_{-i\infty}^{i\infty}
\frac{{\rm d} p_0}{4\pi i}\left[f(p_0)+f(-p_0)\right]}}
{{ +
\int_{-i\infty+\delta}^{i\infty+\delta}
\frac{{\rm d} p_0}{2\pi i}
\frac{f(p_0)+f(-p_0)}{{\rm e}^{p_0/T}-1}}}
\ee
in the two sums of Eq. (\ref{1fe2}) and closing the contour 
on the right side of the complex plane where $1/p^2$ and $1/q^2$ have poles at 
$p_0=|\vec p|$ and $q_0=|\vec q|$, the second order result for the
free energy can be expressed a
\be\label{1fe3}
\tilde\Omega^{(2)}(T,\theta) = e^2\,(d-2)^2
\int\frac{{\rm d}^{d-1}\, q}{(2\pi)^{d-1}}
\int\frac{{\rm d}^{d-1}\, p}{(2\pi)^{d-1}}
\frac{N_B(|\vec q|)}{|\vec q|} 
\frac{N_B(|\vec p|)}{|\vec p|}
\sin^2\left(\frac{p\times q}{2}\right),
\ee
where $N_B$ is given by Eq. (\ref{1defbose1}).
This result already takes into account that
pieces containing less than two Bose distributions vanish in
dimensional regularization \cite{Brandt:2001ud}. 
Of course there is no need to keep an arbitrary $d$ in Eq. (\ref{1fe3}) and
in what follows we will consider $d=4$.
Even without explicitly performing the integral we
notice that the two-loop contribution in Eq. (\ref{1fe3}) is positive 
while the free theory gives the negative value in Eq. (\ref{free01}).

\begin{figure}[h!]
\begin{center}
\includegraphics[scale=0.6]{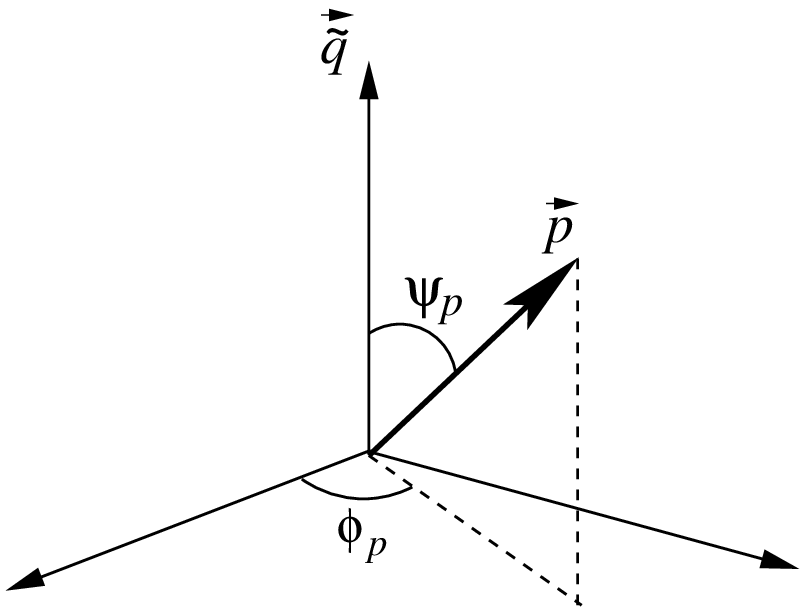}\;\;\;\qquad\includegraphics[scale=0.6]{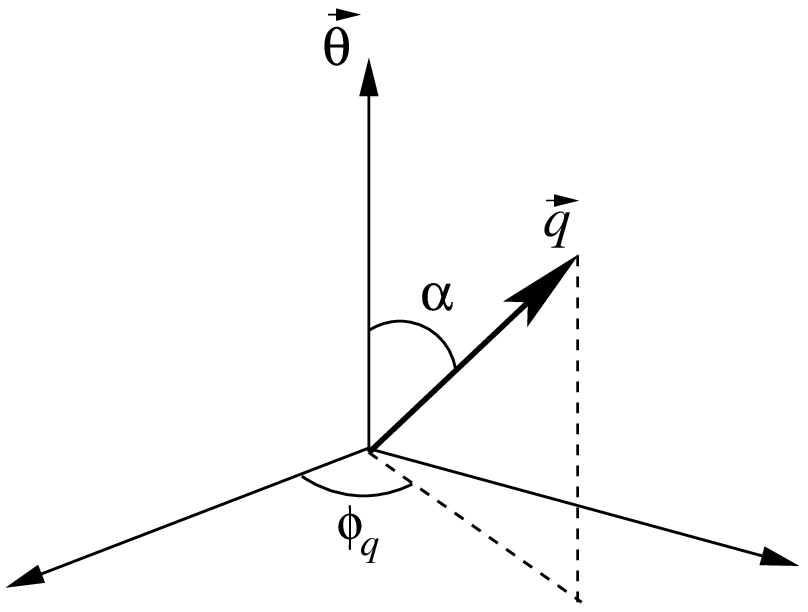}
\end{center}
\caption{}\label{1sysP}
\end{figure}

Let us compute the two-loop integrals in Eq. (\ref{1fe3}).
Introducing the quantity $\vec \theta$ such that
\begin{equation}\label{1eq9}
\epsilon_{ijk}\,\theta_k = \theta_{ij}
\end{equation}
where $\epsilon_{ijk}$ is the Levi-Civita symbol
(notice that 
$\theta_i \equiv \frac 1 2 \epsilon_{ijk}\, \theta_{jk}$),
and employing the integration variables as depicted in figure 
\ref{1sysP}, with $\vec{\tilde q} \equiv \vec q\wedge\vec\theta$, as well as the relation
\begin{eqnarray}\label{1eq8}
p \times q & =  &  p\cdot \tilde q = |\vec p|\, |\vec{\tilde q}|\,\cos\psi_{p} 
= |\vec p|\, |\vec{q}\wedge \vec\theta|\, \cos\psi_{p} 
\nonumber \\ 
& = &
|\vec p|\,|\vec q|\,|\vec\theta|\,\sin\psi_{q}\,\cos\psi_{p},
\end{eqnarray}
and performing the elementary $\psi_p$, $\phi_p$ and $\phi_q$ integrations 
Eq. (\ref{1fe3}) and using (\ref{1boseint1}) yields
\be\label{1fe4}
\tilde\Omega^{(2)}(T,\theta) = 
\frac{e^2\, T^4}{72} - \frac{e^2\, T^4}{2\,\pi^4}\frac{J(\tau)}{\tau},
\ee
where we have introduced the function 
\be\label{1Jtau}
J(\tau) = \frac 1 2 \int_0^\infty {\rm d} u \, \int_0^\infty {\rm d}v 
\int_0^\pi {\rm d} \alpha
\frac{\sin\left(\tau\, u\, v\, \sin(\alpha)\right)}{(\exp{(u)} - 1)(\exp{(v)} - 1)}.
\ee
Performing the angular integration, which can be represented in terms of the Struve
function $H_{0}$ \cite{gradshteyn}, we obtain
\be\label{1Jtau1}
J(\tau) = \frac{\pi}{2}\, \int_0^\infty {\rm d} u \, \int_0^\infty {\rm d}v 
\frac{H_{0}(\tau\, u\, v)}{(\exp{(u)} - 1)(\exp{(v)} - 1)}.
\ee
An equivalent form of this result can also be found in reference \cite{Arcioni:1999hw}
expressed in terms of a four dimensional integral. 

For small values of $\tau$ Eq. (\ref{1Jtau}) can be easily computed by
Taylor expanding $\sin\left(\tau\, u\, v\, \sin(\alpha)\right)$. Then,
at any order in $\tau$ the angular integration is trivial and the
integrations over $u$ and $v$ can be done using Eq. (\ref{1boseint1}).  
The leading term of this expansion will cancel the first term
in Eq. (\ref{1fe4}) and the first non-vanishing term is
\be\label{1fe4a}
\tilde\Omega^{(2)}(T,\theta) \approx \frac{e^2\, T^4}{2}
\left(\frac{\pi^2}{45}\right)^2 {\tau^2}.
\ee
This result is in agreement with reference \cite{Arcioni:1999hw} if we
take into account that they use a definition of $\theta$ which is
twice ours.

\begin{figure}[h!]
\includegraphics[scale=0.3]{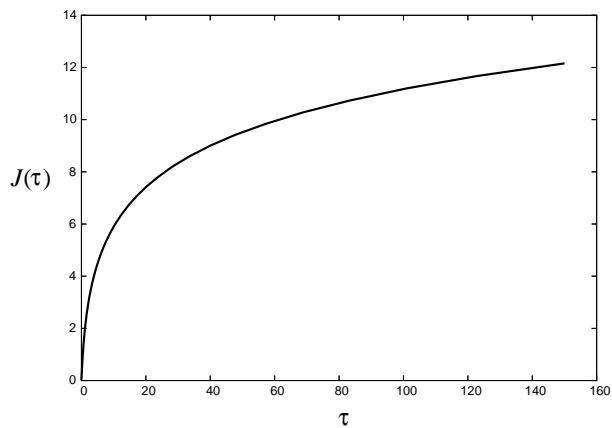}
\caption{The numerical plot of the function given by Eq. (\ref{1Jtau1}).} \label{numint_new}
\end{figure}

From Eq. (\ref{1Jtau}) one can also perform a numerical integration and obtain
the functional dependence for any value of $\tau$. 
The figure \eqref{numint_new} shows a plot of the function $J(\tau)$ which
clearly indicates that $J(\tau)/\tau$ vanishes for large values of $\tau$.  
This property can also be verified using the asymptotic behavior of
the Struve function for large values of its argument which yields 
\be
J(\tau) \rightarrow \left(\frac{\pi}{2}\right)^2\log\tau.
\ee
Since $J(\tau)$ grows logarithmically, the high temperature behavior of
the two loop result, given by Eq. (\ref{1fe4}), converges to a $\tau$
independent result. It is remarkable that the resulting large $\tau$
behavior, given by the first term in Eq. (\ref{1fe4}),
coincides with the two-loop SU(2) free-energy 
modulo a factor of three which counts the internal degrees of freedom
associated with the SU(2) vector fields \cite{Kapusta:1979fh,Arnold:1994ps}. 
This correspondence between large $\tau$ and SU(2)
has been found before in the computation of other thermal
Green functions in noncommutative QED \cite{Brandt:2002aa}.
Previously it has been found that in the regime of large $\tau$, 
or $\sqrt{\theta} \gg 1/T$,
the non planar sector of scalar fields has a thermodynamics resembling
that of a \hbox{$d=2$} dimensional field theory \cite{Fischler:2000fv}.
In the present context of NCQED, the large $\tau$ behavior 
is consistent with the finite infrared behavior of the two loop contribution in
commutative Yang-Mills theory. Indeed, since the two-loop contribution 
to the SU(2) free-energy is finite, it is natural that the corresponding
noncommutative QED result is independent of $\tau$ for large $\tau$. 
At three loop order it is known that there are IR divergences in the 
Yang-Mills free energy. This suggests that the higher order
contributions to the NCQED free energy may be dependent on $\tau$.
In the following section we will consider the three loop contributions.

\newpage

\section{The three-loops contributions}

\begin{figure} 
\[
\begin{array}{cc}
\frac{1}{8}\;\begin{array}{c}\includegraphics[scale=0.6]{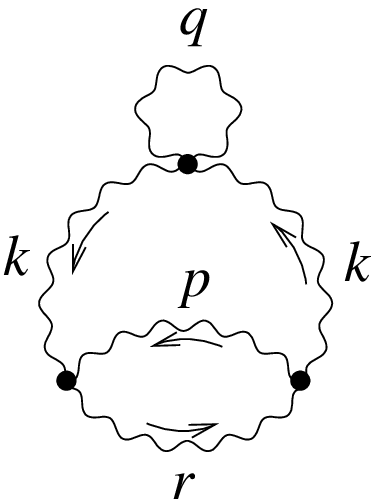}\end{array}&
\frac{1}{16}\;\begin{array}{c}\includegraphics[scale=0.6]{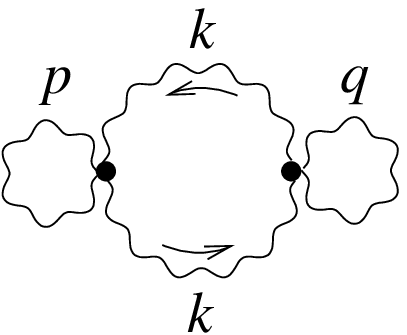}\end{array} \\
    &     \\
(a) & (b) \\
    &     \\
\frac{1}{16}\;\begin{array}{c}\includegraphics[scale=0.6]{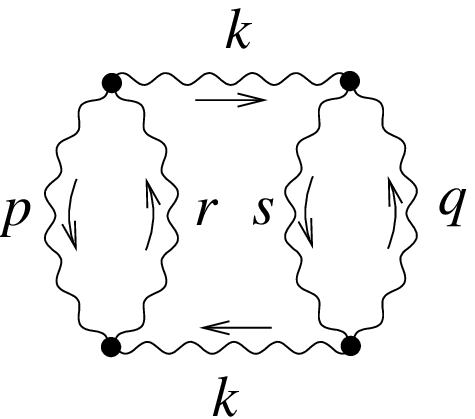}\end{array}& 
-\frac{1}{4}\;\begin{array}{c}\includegraphics[scale=0.6]{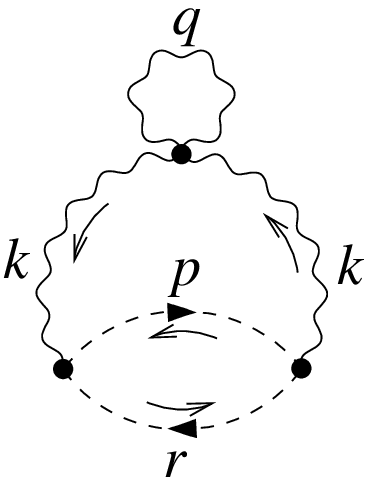}\end{array} \\
    &     \\
(c) & (d) \\
    &     \\
\frac{1}{4}\;\begin{array}{c}\includegraphics[scale=0.6]{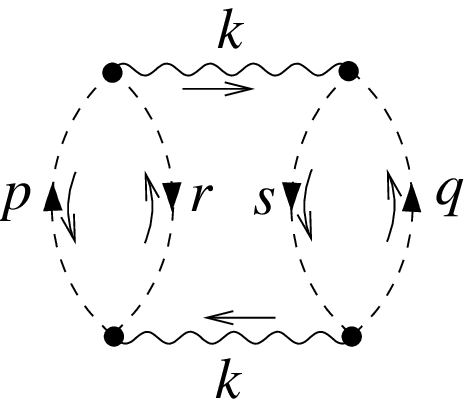}\end{array}&
-\frac{1}{4}\;\begin{array}{c}\includegraphics[scale=0.6]{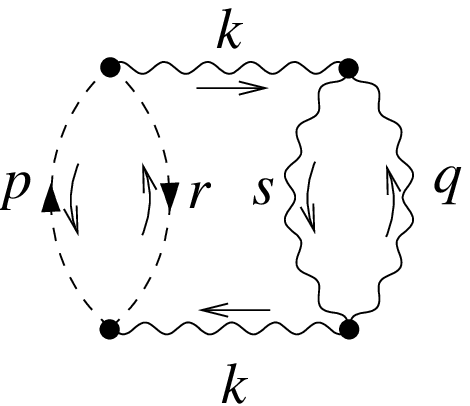}\end{array} \\
    &     \\
(e) & (f) \\
    &     \\
\end{array}
\]
\caption{}\label{g3l1}
\end{figure}

\begin{figure}
\[
\begin{array}{cc}
-\frac{1}{2}\;\begin{array}{c}\includegraphics[scale=0.6]{T4G2_3loops}\end{array}&
\frac{1}{8}\;\begin{array}{c}\includegraphics[scale=0.6]{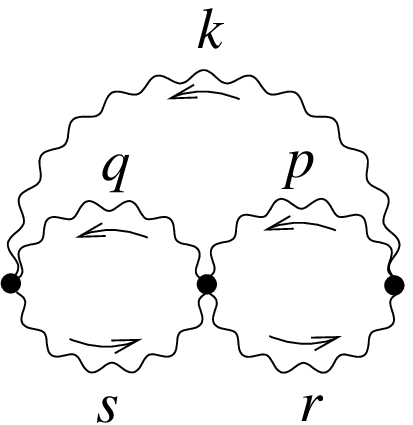}\end{array} \\
    &     \\
(g) & (h) \\
    &     \\
\frac{1}{48}\;\begin{array}{c}\includegraphics[scale=0.6]{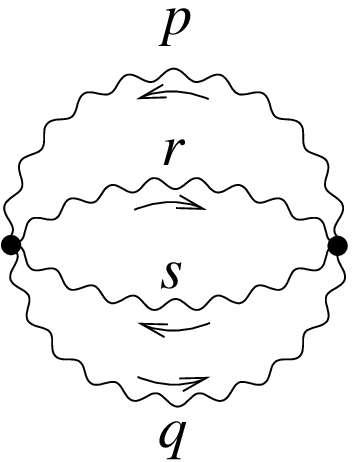}\end{array} &
-\frac{1}{4}\;\begin{array}{c}\includegraphics[scale=0.6]{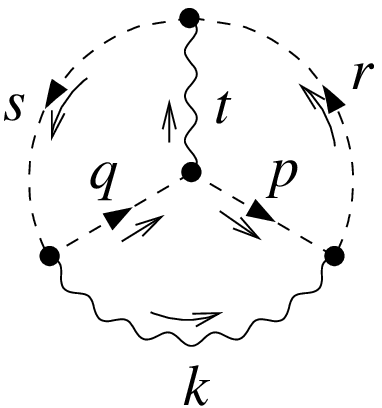}\end{array} \\
\\ &  \\
(i) & (j) 
\\ &  \\
-\frac{1}{3}\;\begin{array}{c}\includegraphics[scale=0.6]{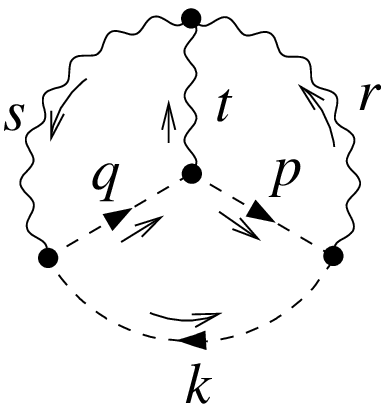}\end{array}&
\frac{1}{24}\;\begin{array}{c}\includegraphics[scale=0.6]{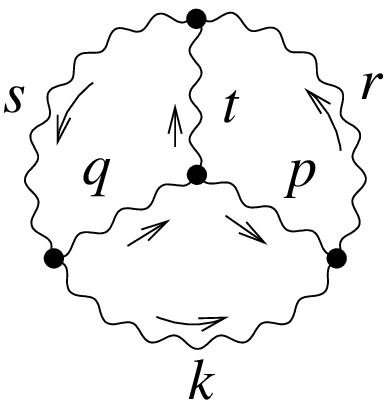}\end{array} \\
    &     \\
  (k) & (l) \\
    &     \\
\end{array}
\]
\caption{}\label{g3l2}
\end{figure}

The three-loop contributions to $\tilde\Omega(T,\theta)$ are shown
in figures \eqref{g3l1} and \eqref{g3l2}.
We have defined $r\equiv k+p$, $s\equiv k+q$ and $t\equiv q-p$ and the arrows
indicate the direction of momenta.
Our basic strategy to deal with these rather involved diagrams consists in first 
reduce the integrands as much as possible using a generalization of
the procedure employed to obtain to Eq. (\ref{1fe2}). In the appendix we
present the details of this rather technical manipulations. Here we
only remark that the shifts in momenta
that can make the integrands simpler are restricted by the more
involved momentum dependence inside the trigonometric factors. Of course
the algebra involved is more complicated than in commutative
gauge theories like QCD \cite{Arnold:1994ps}.

Let us first consider the diagrams in figure \eqref{g3l1}. 
Using the expressions \eqref{Ampa} to \eqref{Ampf}
these contributions combine into the following expression
\begin{eqnarray}\label{1ncqedtot}
\frac{1}{4}
\begin{array}{c}\includegraphics[scale=0.5]{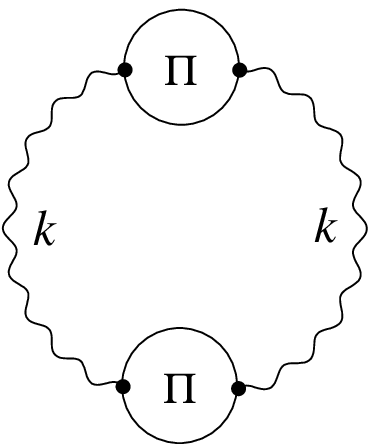}\end{array}
=e^4\tris\left[
16\, {\cal T}_1
\left\{
\frac{1}{4}(d-4)(d-2)^2\,\frac{1}{k^4\,p^2\,q^2} \right.\right.
\nonumber \\
+(d-2)^2 \, \frac{({p\cdot q})^2}{k^4\,p^2\,q^2\,r^2\,s^2}
\nonumber \\
-(d-2)^2\,\frac{1}{k^2\,p^2\,q^2\,r^2}
\nonumber \\ \left.\left.
-\left[\frac{(d-2)^2}{16}-\frac{d}{2}\right]\frac{1}{p^2\,q^2\,r^2\,s^2}
\right\}\right],
\nonumber \\
\end{eqnarray}
where
\[
\tris \equiv
T^3\displaystyle{\int \frac{{\rm d}^{d-1} k}{(2\pi)^{d-1}}   
                 \int \frac{{\rm d}^{d-1} p}{(2\pi)^{d-1}}   
                 \int \frac{{\rm d}^{d-1} q}{(2\pi)^{d-1}}   
\sum_{k_4,p_4,q_4}} 
\]
is a compact notation for the sums and integrations over the three
independent momenta and
\be
{\cal T}_1 \equiv     \sin^2\left({\frac {p\times k}{2}}\right)
                      \sin^2\left({\frac {q\times k}{2}}\right).
\ee
In the left-hand side of  Eq. (\ref{1ncqedtot}) we are using 
a compact representation of the graphs in figure 
\eqref{g3l1} in terms of two insertions
of the self-energy, shown in figure \eqref{ncloop}, with the proper
symmetry factor. There are interesting consequences of the relation 
between the free energy and the self-energy which will be explored
later in this section and in the conclusion. Here we only point out that
the well known problems of infrared divergences 
associated with these so-called {\it ring diagrams} are regulated in
the noncommutative theory by the presence of the factor 
${\cal T}_1$. 

Let us now consider the graphs of figure \eqref{g3l2}.
Before taking into account all these contributions 
let us first focus those terms which are proportional to 
\be
\frac{{\cal T}_1}{p^2\,q^2\,r^2\,s^2}
\ee
as in the last line of Eq. (\ref{1ncqedtot}). The simplest example of
such contributions is the one from {the graph (g)}. Using the Feynman
rules given in Eqs. (\ref{1f1}) and (\ref{1f2}) we readily obtain 
\begin{eqnarray}\label{1simp1}
-\frac{1}{2}
\begin{array}{c}\includegraphics[scale=0.5]{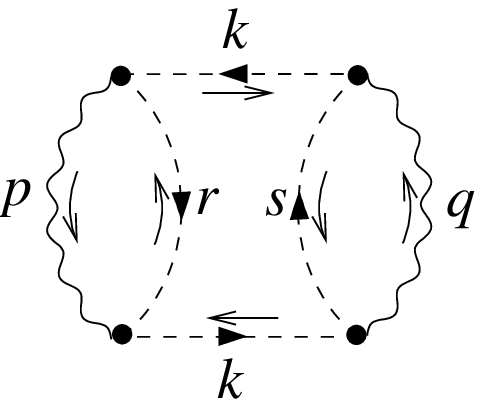}\end{array}
=-e^4\tris\left[
2\,
\frac{{\cal T}_1}{p^2\,q^2\,r^2\,s^2\,k^4}
\left ({k}^{2}+{r}^{2}-{p}^{2}\right )\left ({k}^{2}+{s}^{2}-{q}^{2}\right )
\right],
\end{eqnarray}
where we have used the kinematic relations (\ref{krel1}).
Performing the shifts $p\leftrightarrow -r$ and $q\leftrightarrow -s$,
which does not alter the trigonometric factors and denominators,
the terms involving ${r}^{2}-{p}^{2}$ and ${s}^{2}-{q}^{2}$ vanish and
the resulting expression simplifies to
\begin{eqnarray}\label{1gloop1}
-\frac{1}{2}
\begin{array}{c}\includegraphics[scale=0.5]{T4G2_3loops}\end{array}
=-e^4 \tris\left[
2\,
\frac{{\cal T}_1}{p^2\,q^2\,r^2\,s^2}\right].
\end{eqnarray}
This kind of simplification is also important in order unveil the
true power counting of each individual momenta. Indeed, the naïve power counting of 
Eq. (\ref{1simp1}) would lead us to conclude that the integrand had a denominator
proportional to $k^4$.

Adding Eq. (\ref{1gloop1}) with
the other two contributions from Eqs. (\ref{Amph}) and (\ref{Ampi}) of the appendix yields
\be\label{1part1}
-e^4 \left(8 d - 2 (d-2) (d-4)\right)
\frac{{\cal T}_1}{p^2\,q^2\,r^2\,s^2}.
\ee
Combining Eq. (\ref{1part1}) with the last line of Eq. (\ref{1ncqedtot}) produces
\be\label{1part2}
e^4 \left[2\,(d-2)(d-4) - (d-2)^2\right]
\frac{{\cal T}_1}{p^2\,q^2\,r^2\,s^2}.
\ee
The sum of the first three lines of the integrand of Eq. (\ref{1ncqedtot}) plus the
previous expression gives the full integrand of the three-loops
contributions which are proportional to ${\cal T}_1$.
As we can see this part of the integrand vanishes for $d=2$.

The remaining contributions from figure \eqref{g3l2} comes from part
of the graphs (h) and (i) as well as the ``mercedes'' graphs 
in (j), (k) and (l). We shown in the appendix, that the trigonometric
factors of these contributions can all be reduced to a single factor
given by
\be
{\cal T}_2=\sin\left({\frac {p\times q}{2}}\right)
         \sin\left({\frac {r\times s}{2}}\right)
         \sin\left({\frac {k\times p}{2}}\right)
         \sin\left({\frac {k\times q}{2}}\right).
\ee
Although a direct application of the Feynman rules can produce
other kinds of trigonometric factors like
\be
{\cal T}_3=\sin\left({\frac {p\times s}{2}}\right)
         \sin\left({\frac {q\times r}{2}}\right)
         \sin\left({\frac {k\times p}{2}}\right)
         \sin\left({\frac {k\times q}{2}}\right),
\ee
there are simple identities like
\be\label{1oddI}
e^4 \tris
\frac{{\cal T}_2-{\cal T}_3}{k^2\,p^2\,r^2\,s^2} = 0
\ee
(to verify this identity one just perform the shift $p\leftrightarrow -r$)
which makes it possible to express all the contributions in terms of
${\cal T}_1$. The final result from these contributions is given by
\be\label{1part3}
e^4 \tris\,{\cal T}_2\left[
\frac{2(d-2) (d+2)}{p^2 q^2 r^2 s^2}
-\frac{8(d-2)^2 p\cdot(q+s)}{k^2 p^2 q^2 r^2 s^2}
\right].
\ee
Notice that the second term in the above expression would vanish in QCD
because trigonometric factor ${\cal T}_2$ would be absent (there would
be a color factor instead)

Combining the first three lines of Eq. (\ref{1ncqedtot}) with
Eqs. (\ref{1part2}) and  (\ref{1part3}) we obtain the following
contribution to the three-loop free energy
\begin{eqnarray}\label{1ncqedtot4}
\tilde \Omega^{(4)}(T,\theta) 
= - e^{4} \tris\left\{
16\, {\cal T}_1
\left[
\frac{1}{4}(d-4)(d-2)^2\,\frac{1}{k^4\,p^2\,q^2} \right.\right.
\nonumber \\
+(d-2)^2 \, \frac{({p\cdot q})^2}{k^4\,p^2\,q^2\,r^2\,s^2}
\nonumber \\
-(d-2)^2\,\frac{1}{k^2\,p^2\,q^2\,r^2}
\nonumber \\ \left.\left.
+\frac{2\,(d-2)(d-4)-(d-2)^2}{16 \, p^2\,q^2\,r^2\,s^2}
\right]\right.+
\nonumber \\ \left.
{\cal T}_2\left[
\frac{2(d-2) (d+2)}{p^2 q^2 r^2 s^2}
-\frac{8(d-2)^2 p\cdot(q+s)}{k^2 p^2 q^2 r^2 s^2}
\right]\right\}.
\end{eqnarray}

From Eq. (\ref{1ncqedtot4}) one can now undertake the more challenging
task of computing the sums and integrals. This is also a well defined
expression which does not have infrared divergences. 
However, one should remember
that, as we have already pointed out, at the order $e^4$ the
renormalization of the coupling constant have to be taken into
consideration. Therefore, for arbitrary values of the temperature the
result would be incomplete. On the other hand, we can investigate some extreme
limits of Eq. (\ref{1ncqedtot4}) which may be gauge invariant and
have a meaning by itself. One such limit is the {\it high temperature}
limit. 

\begin{figure}[h!]
\begin{center}
\includegraphics[scale=0.6]{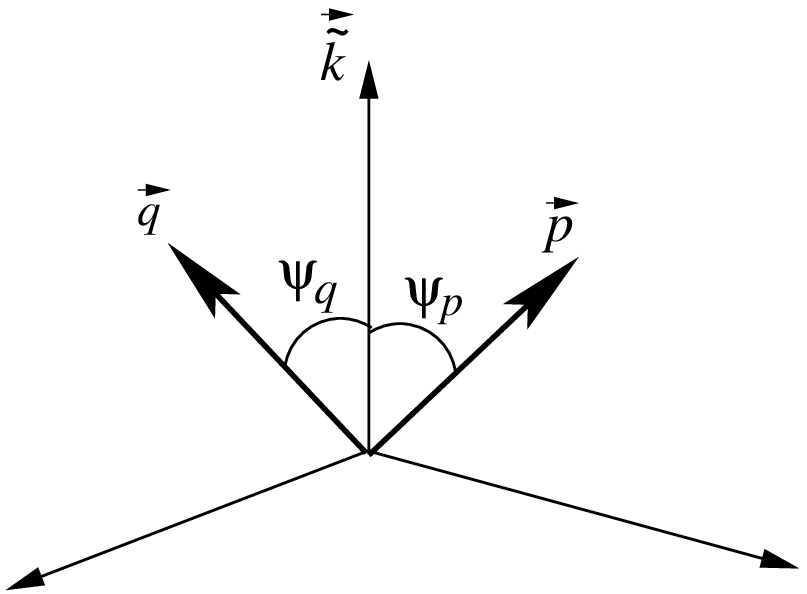}\;\;\;\qquad\includegraphics[scale=0.6]{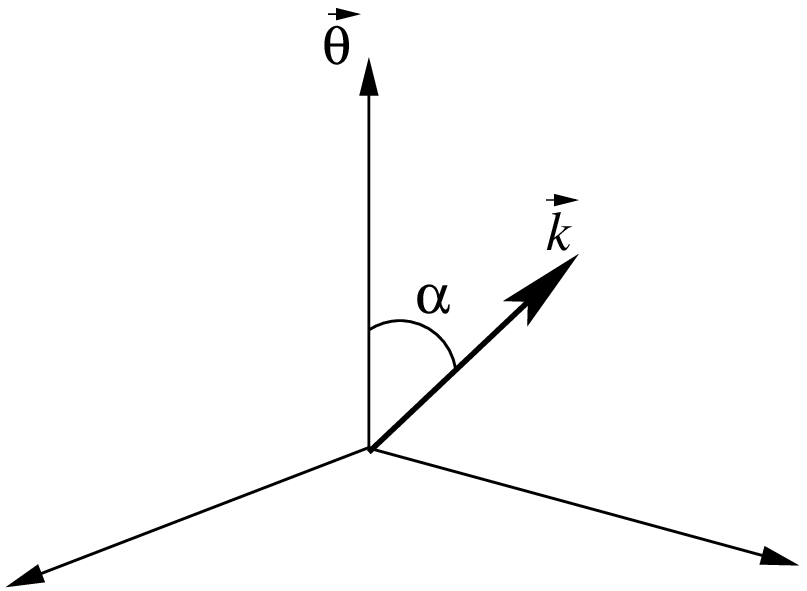}
\end{center}
\caption{}\label{1sys3la}
\end{figure}

In noncommutative field theory the temperature is certainly high
when $T\gg 1/\sqrt{\theta}$ or, equivalently, $\tau\gg 1$. 
In order to investigate the regime of large $\tau$ it is more appropriate
to perform the following rescalings
\begin{eqnarray}\label{defmom}
(p_4,\vec p) & \rightarrow &  T\,(  p_4, \vec{ p}),
\nonumber \\
(q_4,\vec q) & \rightarrow &  T\,(  q_4, \vec{ q})\;\; {\rm and} 
\nonumber \\
(k_4,\vec k) & \rightarrow &  T\,\left(k_4, {\vec { k}}/{\tau}\right), 
\end{eqnarray}
where the momenta variables on the right-hand side are dimensionless.
One obvious advantage of the new dimensionless momentum variables is that the
trigonometric factor ${\cal T}_1$ becomes independent of $\tau$.
Indeed, using the variables indicated in the figure \eqref{1sys3la} 
(these are the only variables which will be relevant in the limit $d=4$ to be
considered later) and performing the rescalings (\ref{defmom}) yields
\be
k\times p = |\vec k| |\vec p| |\vec\theta| \sin\alpha \cos\psi_p
\rightarrow |\vec{ k}| |\vec{ p}| \sin\alpha \cos\psi_p,
\ee
and
\be
k\times q = |\vec k| |\vec q| |\vec\theta| \sin\alpha \cos\psi_q
\rightarrow |\vec{ k}| |\vec{ q}| \sin\alpha \cos\psi_q
\ee
which, in turn, make ${\cal T}_1$ independent of $\tau$.
In terms of the new momentum variables the dependence on $\tau$ is transfered to the
denominators.  Then the terms containing
\be
\frac{1}{k^4} = \frac{1}{[(2\pi n)^2+|\vec{ k}|^2/\tau^2]^2}
\ee
yield a dominant  $\tau^4$ behavior when $n=0$ (the zero mode). 
From this simple analysis we see that the first two terms in Eq. (\ref{1ncqedtot4}) 
are dominant when $\tau$ is large.
Taking into account that, under the rescaling (\ref{defmom}), the measure
transforms as ${\rm d}^{d-1} k \rightarrow T^{d-1} {\rm d}^{d-1} k/\tau^{d-1}$, 
the first two terms in Eq. (\ref{1ncqedtot4}) yield a leading
contribution 
which can be written as
\begin{eqnarray}\label{1ncqedtot4a}
\tilde \Omega^{(4)}(T,\theta) \approx
- 16\,e^{4} (d-2)^2\, T^d \, \tau^{5-d}
\displaystyle{\int        \frac{{\rm d}^{d-1} k}{(2\pi)^{d-1}} 
\frac{1}{|\vec k|^4} \int \frac{{\rm d}^{d-1} p}{(2\pi)^{d-1}}   
                     \int \frac{{\rm d}^{d-1} q}{(2\pi)^{d-1}}   
{\cal T}_1} \nonumber \\
\displaystyle{\sum_{p_4,q_4}
\left(
\frac{1}{4}\frac{d-4}{p^2\,q^2} +  \frac{({p\cdot q})^2}{p^4\,q^4}
\right)},
\end{eqnarray}
where all the integration variables are dimensionless.
Notice that the factor $\tau^{5-d}$ can be viewed
as an infrared regularization which would not be present in QCD.

There are some interesting features about the result in Eq. (\ref{1ncqedtot4a}). 
First, we notice that it has been entirely generated from the ring contribution
given in Eq. (\ref{1ncqedtot}). Therefore we should be able to
reproduce the structure of the integrand in Eq. (\ref{1ncqedtot4a})
from a direct calculation of the following quantity 
\be
{\rm Tr} \,\tilde \Pi(p,k) \,\tilde \Pi(q,k)  = 
                              \tilde\Pi_{\mu}^{\;\;\nu}(p,k)\,
                              \tilde\Pi_{\nu}^{\;\;\mu}(q,k),
\ee
where $\tilde\Pi_{\mu\nu}(p,k)$ is the integrand of the photon self-energy,
in the limit of large $\tau$, or, equivalently setting
$k_4=|\vec k|=0$ except inside the trigonometric factors.
With this prescription, 
the calculation of the graphs shown in figure \eqref{ncloop} gives
\be\label{1tildePi}
\tilde\Pi_{\mu\nu}(p,k) =
4\, e^2\,(d-2)\, \sin^2\left(\frac{k\times p}{2} \right)
\left[\frac{2 p_\mu p_\nu}{p^4} - \frac{\eta_{\mu\nu}}{p^2}\right].
\ee
Then it is easy to verify that
\be
-\frac{1}{4}
{\rm Tr} \,\tilde \Pi(p,k) \,\tilde \Pi(q,k)  = 
-16\,e^4\,(d-2)^2\, {\cal T}_1 
\left[
\frac{1}{4}\frac{d-4}{p^2\,q^2} +  \frac{({p\cdot q})^2}{p^4\,q^4}
\right].
\ee
Comparing this expression with Eq. (\ref{1ncqedtot4a}) we obtain
\be\label{1compact1}
\tilde \Omega^{(4)}(T,\theta) 
\approx -\frac{1}{4} \,T^d \, \tau^{5-d}
\displaystyle{\int        \frac{{\rm d}^{d-1} k}{(2\pi)^{d-1}}}
\frac{1}{|\vec k|^4} 
\Pi_{\mu}^{\;\;\nu}(k) \, \Pi_{\nu}^{\;\;\mu}(k) ,
\ee
where
\be\label{1selfhighT}
\Pi_{\mu\nu}(k) = \int \frac{{\rm d}^{d-1} p}{(2\pi)^{d-1}}   
\sum_{p_4} \tilde\Pi_{\mu\nu}(p,k).
\ee

\begin{center}
\begin{figure} 
\[
\begin{array}{ccc}
\displaystyle{-}
\begin{array}{c}\includegraphics[scale=0.6]{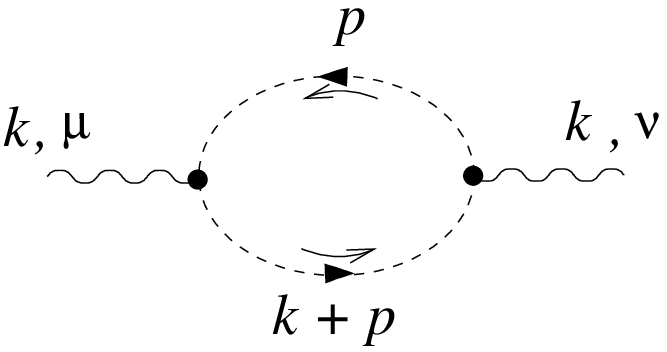}\end{array}&
\displaystyle{\frac{1}{2}}\;\begin{array}{c}\includegraphics[scale=0.6]{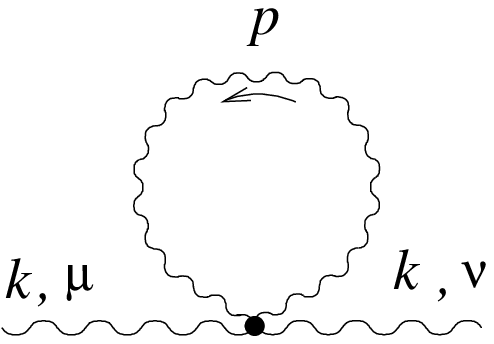}\end{array}&
\displaystyle{\frac{1}{2}}\;\begin{array}{c}\includegraphics[scale=0.6]{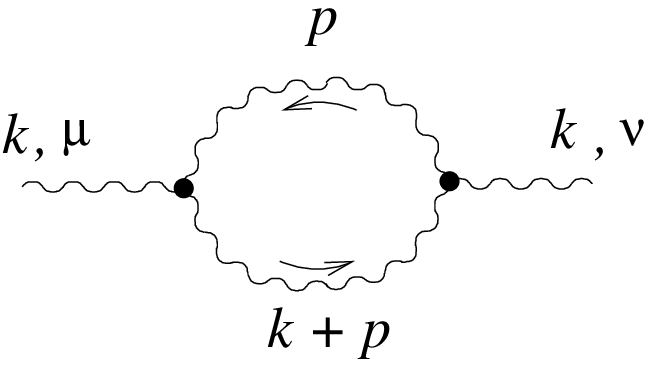}\end{array}
\\  
(a) & (b) & (c)
\end{array}
\]
\caption{}\label{ncloop}
\end{figure}
\end{center}

Another important property of the high temperature limit of the
three-loop free energy as expressed in the form Eq. (\ref{1compact1}), 
is that it may be viewed as the first term of a sequence made by successive
insertions of the self-energy. As we have already pointed out 
this series of {\it ring diagrams} is free of infrared
divergences. Nevertheless, we note that each successive
term in this series will have (for $d=4$) an extra power of $(e\tau)^2$, since
each self-energy contributes with a factor $e^2$ and a factor
$\tau^2$ associated with the extra photon denominator.
Form this point of view we see a breakdown of the formal perturbative
series in the limit when $\tau\gg 1$ ($T\gg 1/\sqrt{\theta}$).
However, as we will see in the next section we can sum the series
of ring diagrams in a closed form and obtain a non-analytic behavior 
in the coupling $e$.

The third property of Eq. (\ref{1compact1}), which also generalizes 
to all higher order ring diagrams, is its gauge independence.
To prove this property we first remark that the result
for $\tilde\Pi_{\mu\nu}$ in Eq. (\ref{1tildePi}) has been obtained
using the general co-variant gauge propagator in Eq. (\ref{1f1})
(the gauge dependent terms are suppressed by powers of $k$). 
In order to complete the proof, let us finish the calculation
of $\Pi_{\mu\nu}$  and shown explicitly
that the contraction of $\Pi_{\mu\nu}$ with the general gauge propagator 
in Eq. (\ref{1f1}) is independent of the gauge parameter $\xi$.
Although the self-energy has been computed previously, we present 
here yet another derivation more akin to our present approach.

From the structure of $\Pi_{\mu\nu}$ given by  Eqs. (\ref{1tildePi})
and (\ref{1selfhighT}) we can write the following general expression 
\be\label{1pi3}
\Pi_{\mu\nu} = \Pi_{00}\,u_\mu u_\nu 
+\Pi_{\rm nc}\, \frac{\tilde k_\mu \tilde k_\nu}{{\tilde k}^2}
+ \Pi_{11}\,\frac{k_\mu k_\nu}{k^2}
             + \Pi_{22}\,\left(\eta_{\mu\nu} - u_\mu u_\nu - 
                                   \frac{k_\mu k_\nu}{k^2} 
 - \frac{\tilde k_\mu \tilde k_\nu}{{\tilde k}^2}\right),
\ee
where $\Pi_{00}$, $\Pi_{\rm nc}$, $\Pi_{11}$ and $\Pi_{22}$ 
are functions of $\tilde k = k \sin\alpha$
(structures which are odd in $k_\mu$ or ${\tilde k}_\mu$ are not compatible
with  the symmetry of $\Pi_{\mu\nu}$). In order to perform the 
integration in Eq. (\ref{1selfhighT}) we are free to choose
$u_\mu=(1,0,\cdots,0)$, $k_\mu=k\,(0,1,\cdots,0)$ and ${\tilde k}_\mu=\tilde k\,(0,0,\cdots,1)$
(in the static limit $k_0=0$).
Then, equating Eqs. (\ref{1selfhighT}) and (\ref{1pi3}) and contracting
with the four tensors in Eq. (\ref{1pi3}) we obtain
\begin{subequations}
\bea\label{1A1}
\Pi_{00} & = & 
-4(d-2)\,e^2\,\int \frac{{\rm d}^{d-1} p}{(2\pi)^{d-1}} 
\sin^2\left(\frac{p\cdot {\tilde k}}{2}\right) 
\sum\left(\frac{1}{p^2} + 2\frac{|\vec p|^2}{p^4} \right),
\\ \label{1D1} 
\Pi_{\rm nc} & = & 
4(d-2)\,e^2\,\int \frac{{\rm d}^{d-1} p}{(2\pi)^{d-1}}  \sin^2\left(\frac{p\cdot {\tilde k}}{2}\right)
\sum\left(\frac{1}{p^2} + 2\frac{(\vec p\cdot\vec {\tilde k})^2}
{|{\tilde {\vec k}|}^2 p^4} \right),
\\ \label{1B1}
\Pi_{11} & = & 
4(d-2)\,e^2\,\int \frac{{\rm d}^{d-1} p}{(2\pi)^{d-1}}  \sin^2\left(\frac{p\cdot {\tilde k}}{2}\right)
\sum\left(\frac{1}{p^2} + 2\frac{(\vec p\cdot\vec k)^2}{|\vec k|^2 p^4} \right)
\;\; {\rm and}\\ \label{1C1} 
\Pi_{22} & = & 
4(d-2)\,e^2\,\int \frac{{\rm d}^{d-1} p}{(2\pi)^{d-1}}  \sin^2\left(\frac{p\cdot {\tilde k}}{2}\right)
\sum\left[\frac{1}{p^2} + \frac{2}{p^4}\left(|\vec p|^2 -
\frac{(\vec p\cdot\vec k)^2}{|\vec k|^2 } - 
\frac{(\vec p\cdot\vec {{\tilde k}})^2}{{|{\tilde {\vec k}|}}^2 } 
\right)\right].
\eea
\end{subequations}

The Matsubara sums in the previous equations can be easily
performed using Eq. (\ref{1TermVac}) and closing the contour
on the right side of the complex plane. The result can be expressed in terms of
the Bose-Einstein thermal distribution, given by Eq. (\ref{1defbose1}), 
and its derivative as follows
\begin{subequations}
\bea\label{1S1}
\sum\,\frac{1}{p^2} &=& - \frac{N_B(|\vec p|)}{|\vec p|} + (T=0) 
\\
\label{1S2}
\sum\,\frac{1}{p^4} &=& \frac{1}{2\,|\vec p|^2}\left(
\frac{N_B(|\vec p|)}{|\vec p|} - N_B^{\prime}(|\vec p|)\right)+(T=0) .
\eea
\end{subequations}
The $(T=0)$ pieces yield a zero result in the dimensionally
regularized momentum integral  \cite{Brandt:2001ud}
so that we may replace the sums
in Eqs. (\ref{1A1}), (\ref{1D1}), (\ref{1B1}) and (\ref{1C1})
by the first terms in Eqs. (\ref{1S1}) and (\ref{1S2}).
In what follows we will consider $d=4$ since the thermal integrals will
not need a regularization.
Using the thermal part of Eqs. \eqref{1S1} and \eqref{1S2}, 
Eq. (\ref{1A1}) yields
\be
\Pi_{00}= \frac{e^2}{\pi^3} \int {\rm d}^3 p \sin^2\left(\frac{p\tilde k\cos\psi_p}{2}\right)
{N_B^{\prime}(p)}=
\frac{2\, e^2}{\pi^2}
\int_0^\infty {\rm d} p\, 
\left(\frac{1}{{\rm e}^p-1}\right)^\prime
\left(
p^2-p\,\frac{\sin({p\tilde k})}{\tilde k}
\right).
\ee
Integrating by parts,
\be
\Pi_{00}= \frac{2\, e^2}{\pi^2}
\int_0^\infty {\rm d} p\, 
\left(\frac{1}{{\rm e}^p-1}\right)
\left(
-2\,p+\frac{\sin({p\tilde k})}{\tilde k}+p\,\cos({p\tilde k})
\right),
\ee
This integral can be done using the formula (\ref{1boseint1})
as well as \cite{gradshteyn}
\begin{subequations}
\bea\label{1grads1}
\int_0^\infty \frac{x^{2m} \sin{b x}}{{\rm e}^x-1} {\rm d} x & = & 
(-1)^m \frac{\partial^{2m}}{\partial b^{2m}}\left[\frac{\pi}{2}
{\rm cth}(\pi b) -\frac{1}{2 b}\right] \\
\label{1grads2}
\int_0^\infty \frac{x^{2m+1} \cos{b x}}{{\rm e}^x-1} {\rm d} x & = & 
(-1)^m \frac{\partial^{2m+1}}{\partial b^{2m+1}}\left[\frac{\pi}{2}
{\rm cth}(\pi b) -\frac{1}{2 b}\right].
\eea
\end{subequations}
The result is
\be\label{1Af1}
\Pi_{00}(\tilde k) = \frac{2\,e^2}{\pi^2}
\left(\frac{\pi^2}{6}+\frac{\pi}{2\tilde k}{\rm cth}(\pi\tilde k)
-\frac{\pi^2}{2}{\rm cth}^2(\pi\tilde k)\right)
\ee
Proceeding similarly for $\Pi_{\rm nc}$, $\Pi_{11}$ and $\Pi_{22}$ 
we obtain
\be\label{1Df1}
\Pi_{\rm nc}(\tilde k) = -\frac{2\,e^2}{\pi^2}
\left(\frac{\pi^2}{2}-\frac{\pi}{2\tilde k}{\rm cth}(\pi\tilde k)
-\frac{\pi^2}{2}{\rm cth}^2(\pi\tilde k)  + \frac{1}{{\tilde k}^2}\right)
\ee
and $\Pi_{11} = \Pi_{22} = 0$ , where $\tilde k = k\, \sin(\alpha)$. 

Since only $\Pi_{00}$ and $\Pi_{\rm nc}$ are nonzero Eq. (\ref{1pi3}) implies that
(recalling that we have chosen
$u_\mu=(1,0,0,0)$, $k_\mu=k\,(0,1,0,0)$ and ${\tilde k}_\mu=\tilde k\,(0,0,0,1)$)
\be\label{1transv9}
\left. k_\mu\,\Pi^{\mu\nu}\right |_{\rm static} = 
\left. k_i \,\Pi^{ij}\right |_{\rm static} = 0.
\ee
From this transversality property, it follows that the gauge parameter
dependence of the photon propagator will cancel in a ring diagram 
containing any number of self-energies. Therefore the sum of 
all the rings is also gauge independent. Notice that the cancellation
of the gauge dependent part of the photon propagator in (\ref{1f1}) 
takes place inside the integral (\ref{1compact1}) which is finite both in the
infrared and ultraviolet regime.

\section{Non-perturbative contributions}
\begin{figure}[h!]
\begin{center}
\[
\begin{array}{cc}
\bar \Pi_{00}
\begin{array}{c}\includegraphics[scale=0.25]{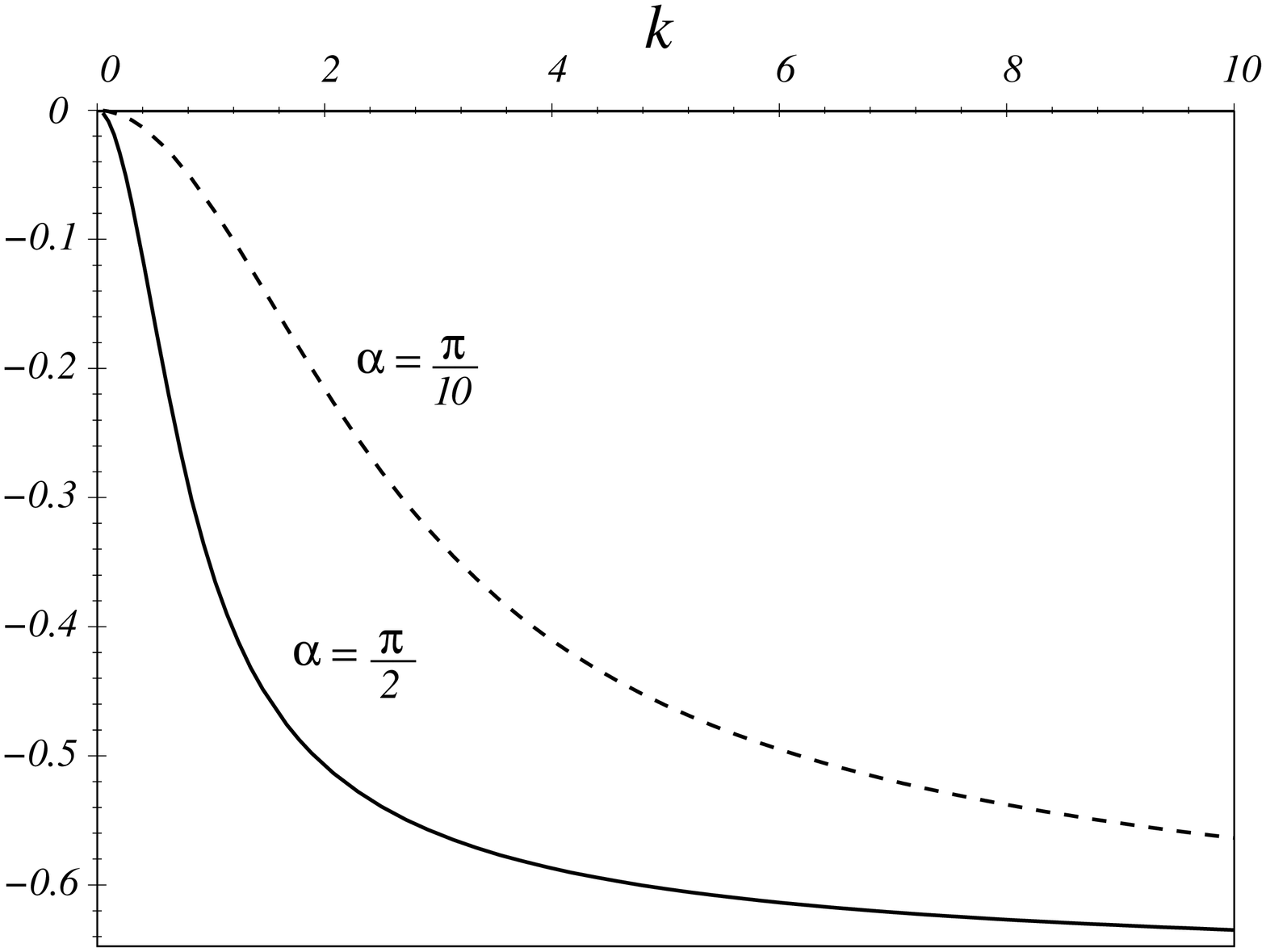}\end{array} &
\;\;\;\;\;\;\bar \Pi_{\rm nc}  
\begin{array}{c}\includegraphics[scale=0.25]{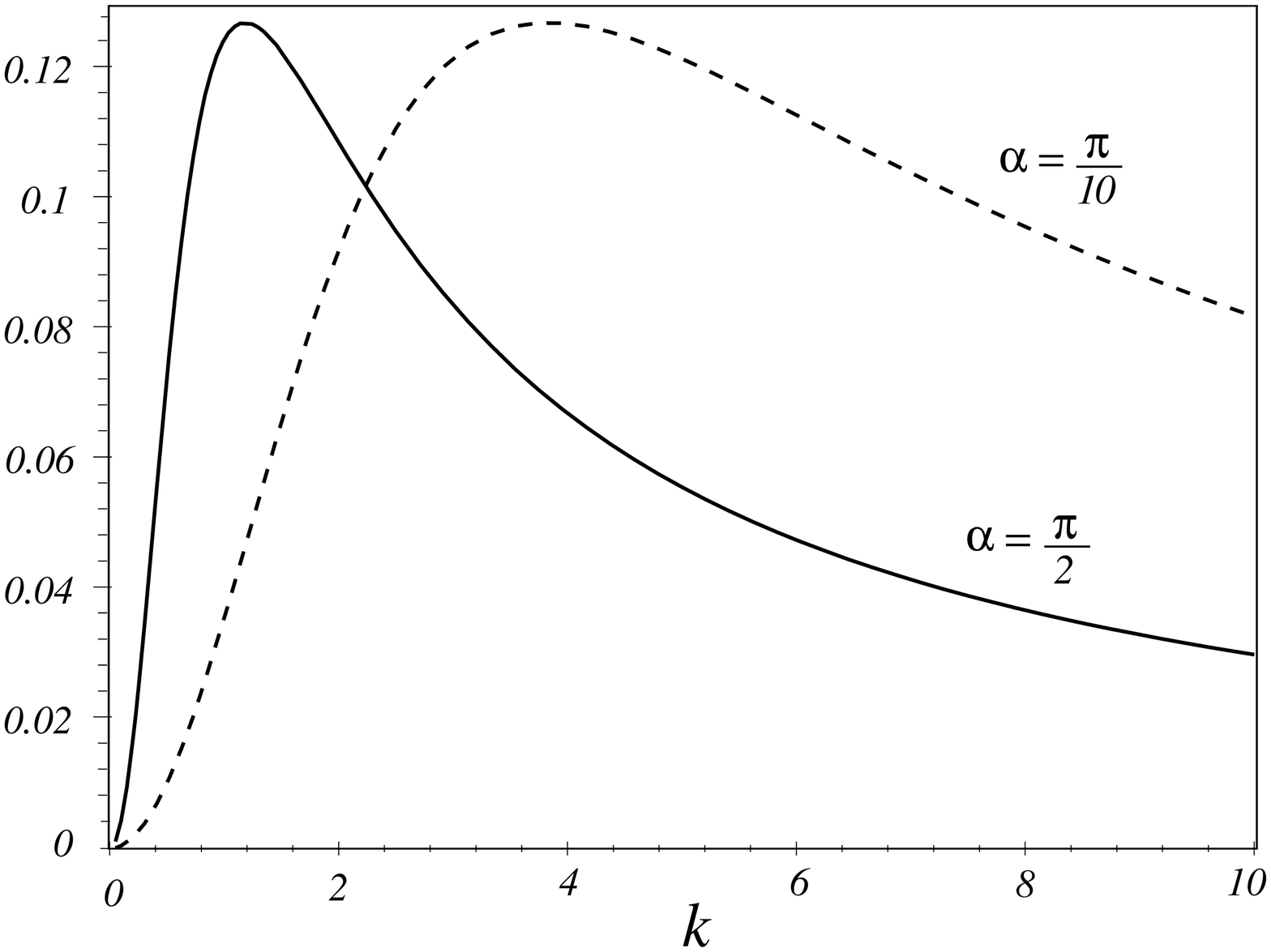}\end{array} \\
\;\;\;\;\;\;\;\;\;\;\;\;\;\;\;\;\; & 
\;\;\;\;\;\;\;\;\;\;\;\;\;\;\;\;\;\;\;\;\;\;\;  \\ &  \\
\;\;\;\;\;\;\;(a)   & \;\;\;\;\;\;\;(b) 
\end{array}
\]
\end{center}
\caption{The longitudinal and transverse modes of the static photon
  self-energy in NCQED.}\label{1AD1}
\end{figure}

Let us now consider the following contribution to the free energy
\be
\tilde\Omega^{\rm ring}(T,\theta) =
-\frac{1}{2}\left[
\frac{1}{2}\begin{array}{c}\includegraphics[scale=0.6]{1PI}\end{array}
-\frac{1}{3}\begin{array}{c}\includegraphics[scale=0.6]{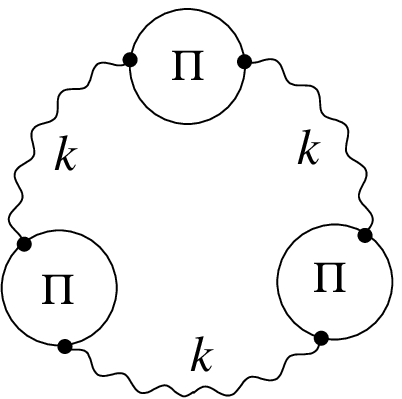}\end{array}
+\cdots ,
\right]
\ee
where the photon propagators are such that $k^2=-|\vec k|^2$.
Taking into account the orthonormality and idempotency of the tensors $u_\mu u_\nu$ 
and $\tilde k_\mu \tilde k_\nu/\tilde k^2$ in Eq. \eqref{1pi3}, as well as 
\be
-\sum_{n=2}^{\infty} \frac{x^n}{n} = x+\log(1-x) 
\ee
we obtain
\begin{eqnarray}\label{1tudo5a}
\Omega^{\rm ring}(T,\theta) =
\frac{1}{2} 
\frac{T^4}{(2\pi\tau )^3}\int{\rm d}^3 k \!\!\!\! \!\!\!\!\! \! & &\left[
\frac{(e\tau)^2\, \bar \Pi_{00}(\tilde k) }{|\vec k|^2} + 
\log\left(1-\frac{(e\tau)^2\, \bar \Pi_{00}(\tilde k) }{|\vec k|^2}\right) 
\right . 
\nonumber \\ \qquad \;\;\;\;\;\;\;\;\;\;\;\;\;\;\;\;\;\;\;\;\;\;\; & + & 
\left.
\frac{(e\tau)^2\, \bar\Pi_{\rm nc}(\tilde k) }{|\vec k|^2} + 
\log\left(1-\frac{(e\tau)^2\, \bar\Pi_{\rm nc}(\tilde k) }{|\vec k|^2}\right) 
\right];\;\;\;\;\; \tilde k = |\vec k| \sin\alpha.
\end{eqnarray}
The functions $\bar \Pi_{00}(\tilde k)$ and $\bar\Pi_{\rm nc}(\tilde k)$ are such that 
$e^2 \bar \Pi_{00}(\tilde k) = \Pi_{00}(\tilde k)$ and  
$e^2 \bar\Pi_{\rm nc}(\tilde k) = \Pi_{\rm nc}(\tilde k)$ where $\Pi_{00}(\tilde k)$ 
and $\Pi_{\rm nc}(\tilde k)$ given by 
Eqs. (\ref{1Af1}) and (\ref{1Df1}) respectively. Performing the 
elementary angular integration this expression can be written as
\begin{eqnarray}\label{1tudo5}
\Omega^{\rm ring}(T,\theta) =
\frac{e^3\,T^4}{(2\pi)^2}\frac{1}{(e\tau)^3}\int_0^{\pi/2}{\rm d}\alpha \sin\alpha
\int_0^{\infty}{\rm d} k \, k^2  
\!\!\!\! \!\!\!\!\! \! & &\left[
\frac{(e\tau)^2\, \bar \Pi_{00}(\tilde k) }{k^2} + 
\log\left(1-\frac{(e\tau)^2\, \bar \Pi_{00}(\tilde k) }{k^2}\right) 
\right . 
\nonumber \\ \qquad \;\;\;\;\;\;\;\;\;\;\;\;\;\;\;\;\;\;\;\;\;\;\; & + & 
\left.
\frac{(e\tau)^2\, \bar\Pi_{\rm nc}(\tilde k) }{k^2} + 
\log\left(1-\frac{(e\tau)^2\, \bar\Pi_{\rm nc}(\tilde k) }{k^2}\right) 
\right]. \nonumber \\ 
\end{eqnarray}
The figure \eqref{1AD1} show the plots of
the functions $\bar \Pi_{00}(\tilde k)$ and $\bar\Pi_{\rm nc}(\tilde k)$.
The asymptotic behavior can be easily obtained from Eqs. 
(\ref{1Af1}) and (\ref{1Df1}). The result for small and large values of
$\tilde k$ are given respectively by
\begin{subequations}\label{1asymp1}
\bea\label{1asymp1b}
\bar \Pi_{00}(\tilde k) & \simeq &-\frac{4\pi^2}{45}{\tilde k^2}
+\frac{4\pi^4}{315}  {\tilde k^4}+\cdots , \\
\label{1asymp1a}
\bar\Pi_{\rm nc}(\tilde k)& \simeq &
\frac{2\pi^2}{45}{\tilde k^2} 
-\frac{8\pi^4}{945}  {\tilde k^4}+\cdots 
\eea
\end{subequations}
and 
\begin{subequations}\label{1asymp1blarge1}
\bea\label{1asymp1blarge}
\lim_{k\rightarrow \infty} \bar \Pi_{00}(\tilde k) & = & 
- \frac{2}{3} \\
\lim_{k\rightarrow \infty}\bar\Pi_{\rm nc}(\tilde k) & = & 0 .
\eea
\end{subequations}

Eq. (\ref{1tudo5}) represents a contribution to the free-energy which
is non-analytic in the coupling constant in the sense that it is not 
a simple power of $e^2$. In order to gain some basic understanding of this
function, let us first consider its asymptotic behavior 
for large values of $e\tau$. In this case it is convenient first
to use integration by parts and the identity
\be
\frac{k^3}{3}\frac{{\rm d}}{{\rm d}k}\left[
\frac{G(k)}{k^2} + \log\left(1-\frac{G(k)}{k^2}\right)\right] =
\frac{1}{3}\frac{G(k)(2\,G(k)-k\,G^\prime(k))}{k^2-G(k)}
\ee
(${}^\prime$ denotes de derivative in relation to $k$).
Taking into account Eqs. \eqref{1asymp1} and \eqref{1asymp1blarge1}
one can easily shown
that the surface term does not contribute and we are left with
\be
\Omega^{\rm ring}(T,\theta) =
-\frac{e^3}{3} \frac{T^4}{(2\pi)^2} (e \tau) \int_0^{\pi/2} 
{\rm d}\alpha \sin\alpha
\int_0^\infty {\rm d} k \left[
\frac{\bar \Pi_{00}(2\,\bar \Pi_{00}-k\,\bar \Pi_{00}^\prime)}
{k^2-{(e\tau)^2\, \bar \Pi_{00}}}
+
\frac{\bar\Pi_{\rm nc}(2\,\bar\Pi_{\rm nc}-k\,\bar\Pi_{\rm nc}^\prime)}
{k^2-{(e\tau)^2\, \bar\Pi_{\rm nc}}}
\right].
\ee
Performing the rescaling $k\rightarrow e\tau k$ 
\begin{eqnarray}
\Omega^{\rm ring}(T,\theta) =
-\frac{e^3}{3} \frac{T^4}{(2\pi)^2 } \int_0^{\pi/2} 
{\rm d}\alpha \sin\alpha
\int_0^\infty {\rm d} k \left[
\frac{\bar \Pi_{00}(e\tau\tilde k)
(2\,\bar \Pi_{00}(e\tau\tilde k)-k\,\bar \Pi_{00}^\prime(e\tau\tilde k))}
{k^2-\bar \Pi_{00}(e\tau\tilde k)} \right. + \nonumber \\
\left.
\frac{\bar\Pi_{\rm nc}(e\tau\tilde k)
(2\,\bar\Pi_{\rm nc}(e\tau\tilde k)-k\,\bar\Pi_{\rm nc}^\prime(e\tau\tilde k))}
{k^2-\bar\Pi_{\rm nc}(e\tau\tilde k)}
\right]. \nonumber \\
\end{eqnarray}
For $e\tau\gg 1$ the integrand is dominated by the asymptotic behavior
of $\bar \Pi_{00}$ and $\bar\Pi_{\rm nc}$ given by the Eq. (\ref{1asymp1blarge1}). Then we can
easily perform the integrals in the previous expression and obtain
\be
\lim_{e\tau\rightarrow\infty} \Omega^{\rm ring}(T,\theta) =
-\frac{e^3}{54\pi}\sqrt{6}\, T^4.
\ee
Comparing the previous expression with the SU(N) result
\cite{Arnold:1994ps}
\be
\Omega^{\rm ring}_{{\rm SU(N)}} = - (N^2-1)\frac{\pi^2\,T^4}{9} 
\frac{16}{\sqrt{3}}\left(\frac{\sqrt{N} e}{4\pi}\right)^3
\ee
we see that, as in the case of the two-loop contribution, the following
relation holds
\be\label{1largeet1}
\lim_{e\tau\rightarrow\infty} \Omega^{\rm ring}(T,\theta) =
\frac{1}{3} \Omega^{\rm ring}_{{\rm SU(2)}}.
\ee

\begin{figure}[h!]
\begin{center}
\[
\begin{array}{cc}
\Re( {{\cal I}(e\tau)})  
\begin{array}{c}\includegraphics[scale=0.25]{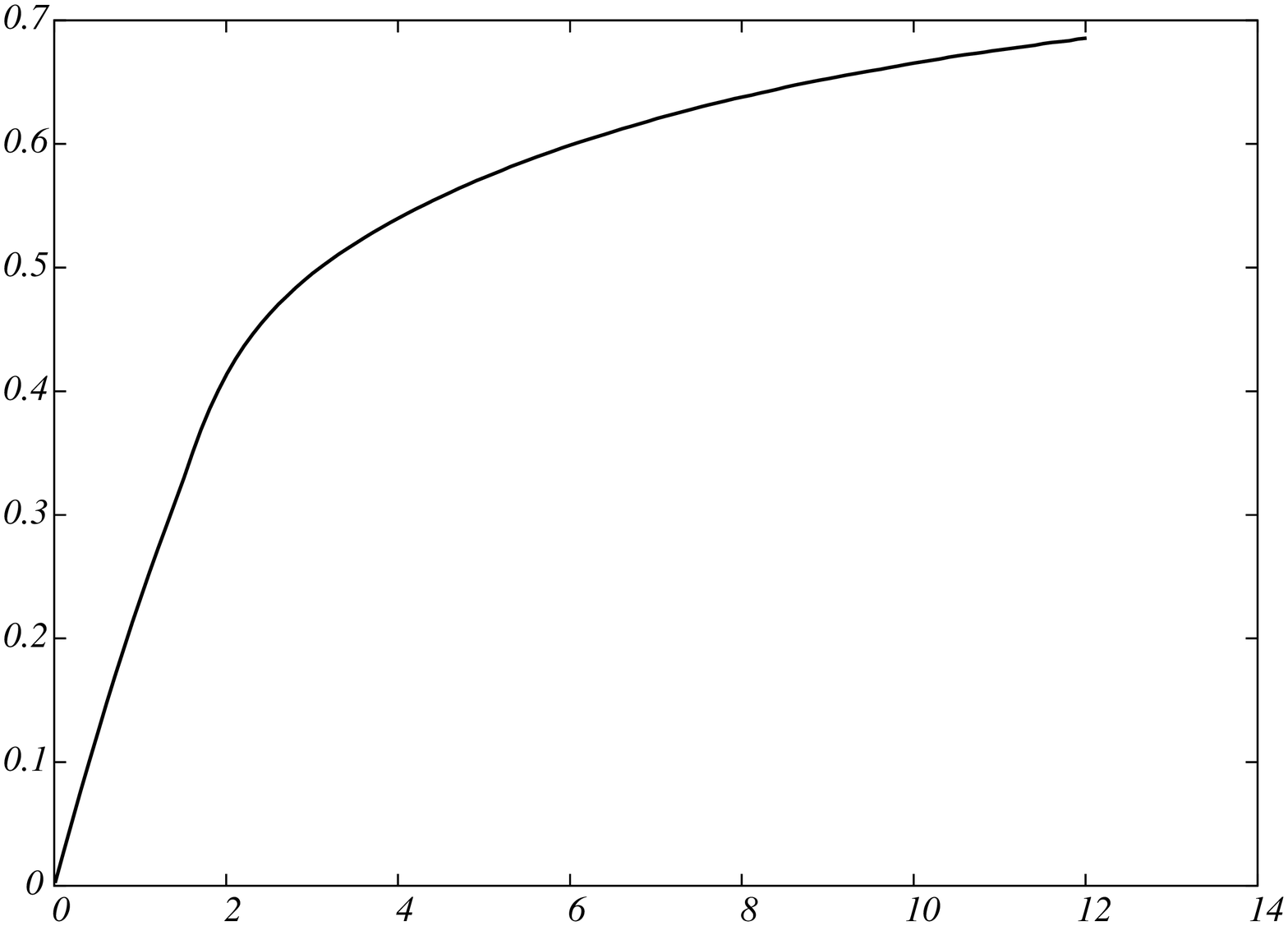}\end{array} &
\;\;\;\;\;\;\Im( {{\cal I}(e\tau)})  
\begin{array}{c}\includegraphics[scale=0.25]{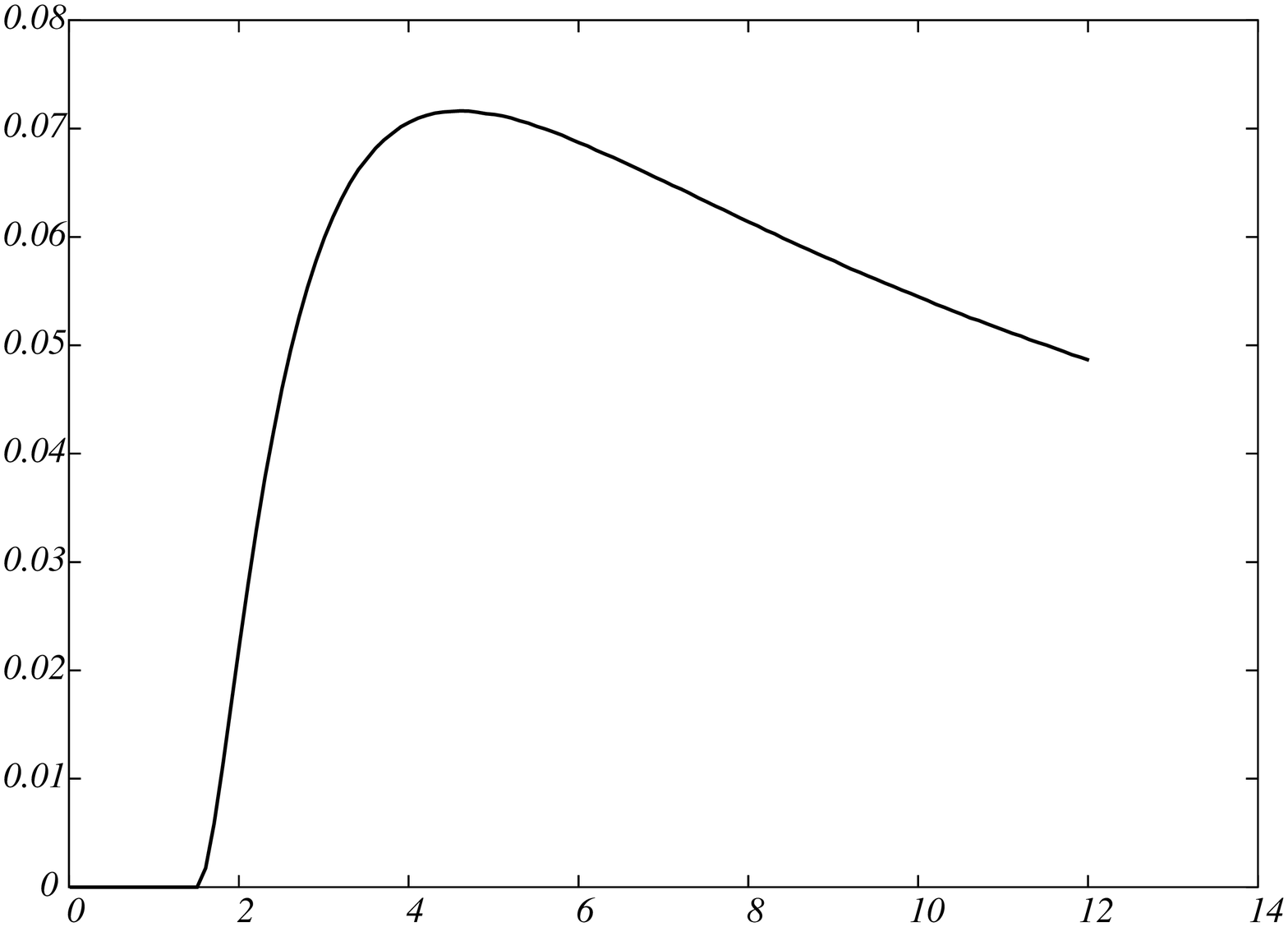}\end{array} \\
\;\;\;\;\;\;\;\;\;\;\;\;\;\;\;\;\;e\tau & 
\;\;\;\;\;\;\;\;\;\;\;\;\;\;\;\;\;\;\;\;\;\;\;e\tau \\ &  \\
\;\;\;\;\;\;\;(a)   & \;\;\;\;\;\;\;(b) 
\end{array}
\]
\end{center}
\caption{
The real (a) and imaginary parts (b) of $\Omega^{\rm ring}$, in units of 
$-\frac{\sqrt{6}}{54\pi}\,e^3 T^4 $.
}\label{1figfe1}
\end{figure}

Let us now analyze the properties of $\Omega^{\rm ring}(T,\theta)$ for
intermediate values of $e\tau$. It is convenient to define the
quantity
\begin{eqnarray}\label{1numint1}
{\cal I}(e \tau) = \frac{27}{12\pi}\frac{\sqrt{6}}{(e\tau)^3} 
\int_0^{\pi/2}{\rm d}\alpha \sin\alpha
\int_0^{\infty}{\rm d} k \, k^2  
\!\!\!\! \!\!\!\!\! \! & &\left[
\frac{(e\tau)^2\, \bar \Pi_{00}(\tilde k) }{k^2} + 
\log\left(1-\frac{(e\tau)^2\, \bar \Pi_{00}(\tilde k) }{k^2}\right) 
\right . 
\nonumber \\ \qquad \;\;\;\;\;\;\;\;\;\;\;\;\;\;\;\;\;\;\;\;\;\;\; & + & 
\left.
\frac{(e\tau)^2\, \bar\Pi_{\rm nc}(\tilde k) }{k^2} + 
\log\left(1-\frac{(e\tau)^2\, \bar\Pi_{\rm nc}(\tilde k) }{k^2}\right) 
\right]. \nonumber \\
\end{eqnarray}
in terms of which we can write Eq. (\ref{1tudo5}) as 
\be\label{1fenum1}
\Omega^{\rm ring}(T,\theta) 
= -\frac{e^3}{54\pi}\sqrt{6}\, T^4 \, {\cal I}(e \tau).
\ee
In the figure \eqref{1figfe1} (a) it is shown a plot of the real part of the
function ${\cal I}(e \tau)$. For relatively small values of $e\tau$ it
grows linearly, as we would expect from expanding the integrand in
Eq. (\ref{1numint1}). In this case Eq. (\ref{1fenum1}) behaves like
$e^4\tau$ which is the three-loop behavior of $\Omega^{\rm ring}(T,\theta)$. 
As $e\tau$ increases, we see from figure \eqref{1figfe1} that the
dependence on $e\tau$ softens and the curve tends to its asymptotic
constant value as we expect from the previous analysis.

We also plot in the figure \eqref{1figfe1} (b) the imaginary part 
of ${\cal I}(e \tau)$. This plot was obtained 
replacing the logarithm in Eq. (\ref{1numint1}) by $-i\pi$ whenever
its argument becomes negative. Notice that only the second logarithm
can have a negative argument, because $\bar\Pi_{\rm nc}$ is always positive
while $\bar \Pi_{00}$ is always negative. In fact we can find the exact
{\it critical value} of $e\tau$ above which the plot in figure 
\ref{1figfe1} (b) becomes non-zero. This can be obtained by imposing
the condition
\be\label{1critical1}
e\tau > \lim_{k\rightarrow 0}\sqrt{\frac{k^2}{\bar\Pi_{\rm nc}}}=\frac{1}{\sin\alpha}
\frac{3}{2\,\pi}\sqrt{10} \ge \frac{3}{2\,\pi}\sqrt{10} ,
\ee
where we have used (\ref{1asymp1a}). 


\section{Discussion}
In this paper we have obtained all the contributions to the free
energy up to the leading non-analytic terms which arises 
from the summation of ring diagrams. In the lowest non-trivial order,
namely the two-loop order, previously investigated in reference \cite{Arcioni:1999hw}, 
our approach allows to cast the results 
in a relatively simple form which gives both the low and high temperature limits 
in a straightforward way. We have also analyzed the three-loop
contributions and identified the leading high temperature
terms. (Three-loop contributions are also relevant to
implement the renormalization scale independence of the free energy, so that a
change in the effective running coupling constant 
$e^2(T)$ is compensated by changes in higher order contributions
starting at ${\cal O}(e^4)$ \cite{Frenkel:1992az}.)
The series of leading higher order contributions, proportional to
powers of $(e\tau)^2$, has been summed exactly and the result was
expressed in terms of a double integral representing the ring
contributions to the free energy as a function of $e\tau$.
Both the two-loop contributions and the sum of the rings have been
computed in a general co-variant gauge and the gauge independence has
been verified for these contributions.

Perhaps the most interesting feature of this analysis is expressed in Eq. (\ref{1critical1}),
which shows that the free energy acquires an imaginary part above a
critical temperature.
Although this has been found from our analysis of the behavior of the free energy,
it is interesting to notice that the critical value 
\be\label{1critical2}
(e\tau)_c = e\,\theta \, T^2_c = \frac{3}{2\,\pi}\sqrt{10}
\ee
can be related with the solution of the equation which gives a
self-consistent gauge independent definition of thermal masses through
the relation \cite{Rebhan:1993az}
\be\label{genthermmass}
m_{A}^2 = - \left. \Pi_{A}(k_0=0,\vec k) 
\right |_{|\vec k|^2 = - m_{A}^2} ,
\ee
where the subindex ``$A$'' in $\Pi_{A}$ represents either ``${00}$'' or ``${\rm nc}$'',
as defined in Eq. (\ref{1pi3}). One can now investigate the solutions
of this equation as a function of the parameter $(e\tau)$. 

Let us first consider the electric ``${00}$'' mode. Using Eq. (\ref{1Af1}) one can easily
find that there is a positive solution for $m_{00}^2$. This solution
occurs for asymptotic values of the parameter $(e\tau)$. Indeed, using
the asymptotic behavior given by Eq. (\ref{1asymp1blarge}) 
and taking into account the inverse of the rescaling transformations in (\ref{defmom}),
we obtain
\be
m_{00}^2 = \frac 2 3 \, e^2 T^2.
\ee
It is interesting to note that this noncommutative Debye mass is
numerically identical to the result of the SU(2) theory \cite{kapusta:book89}.
This is also a consequence of the correspondence between the large
$e\tau$ regime and the commutative non-Abelian gauge theory, as already manifested
in Eq. (\ref{1largeet1}). (We also note that
a naïve definition of the thermal mass such that
$m_{A}^2 = - \lim_{k\rightarrow 0}\Pi_{A}(k_0=0,\vec k )$ 
would be zero in the pure gauge sector of NCQED, because both 
$\bar \Pi_{00}$ and $\bar\Pi_{\rm nc}$ vanish as $k\rightarrow 0$, 
for fixed values of $\tau$. However, in higher orders, such a
mass would be gauge dependent \cite{kapusta:book89}.)

Proceeding similarly with the ``$nc$'' mode, we find that Eq. (\ref{genthermmass})  
admits negative solutions for $m^2_{\rm nc}$. This happens in the regime $e\tau>(e\tau)_c$
which is when the argument of the logarithm in Eq. (\ref{1tudo5a}) becomes imaginary.
For arbitrary values of $e\tau$ one would have to solve Eq. (\ref{genthermmass}) numerically. 
However, when $e\tau$ is close to its critical value, we can obtain the
analytic solution of Eq. (\ref{genthermmass}) using  
the expression given by Eq. (\ref{1asymp1a}), multiplied by $e^2$. Performing a
simple calculation, the nontrivial solution for the mass can be written as
\be
m_{\rm nc}^2 \simeq \frac{21}{4\pi^2}
\left[\left(\frac{(e\tau)_c}{(e\tau)}\right)^2-1\right]
\frac{e^2\,T^2}{(e\tau)_c^2}
\simeq \frac{7 e^2}{30} \left(T_c^2-T^2\right),
\ee 
where $T_c^2$ is defined in Eq. (\ref{1critical2})
(again, we have taken into account the inverse of the rescaling
transformations in (\ref{defmom})).
This solution shows explicitly that $m_{\rm nc}^2$ becomes negative
when  $T>T_c$. 

The contrasting behavior of the two modes is a direct consequence of the fact that 
while $\bar \Pi_{00}$ is always negative, $\bar \Pi_{\rm nc}$ is always positive
(see figure \eqref{1AD1}).
It is remarkable that despite all the similarities between NCQED and
commutative Yang-Mills theories there is such an important difference
as far as the stability is concerned. Of course the important feature
here is the presence of the non-zero static transverse mode 
$\Pi_{\rm nc}$, which would vanish in theories like QCD.
This is consistent with previous findings from the analysis of the 
dispersion relations in noncommutative SYM theories at finite 
temperature \cite{Landsteiner:2001ky}.

In conclusion, we have shown that above a certain critical temperature,
the system may undergo a phase transition which is induced by the 
noncommutative magnetic mode. This behavior will also occur even in the presence
of fermions, because the contributions from the fermion loops are the same
as in the commutative theory. Since in commutative QED the static magnetic
mode is absent and the magnetic mass vanishes to all 
orders \cite{kapusta:book89},
it follows that the fermions will not modify the behavior of the magnetic
mode which is responsible for the instability of the system. It is 
interesting to note that the negative value of the squared magnetic mass
$m_{\rm nc}^2$ is reminiscent of the Jeans mass $M^2 \sim - G T^4$ 
which arises in thermal quantum 
gravity \cite{Gross:1982cv}. Such a mass leads to the apearence of an
imaginary part in the free energy,
which may be related to the decay rate of the quantum metastable vacuum.

\bigskip

\noindent{\bf Acknowledgment}

This work was supported by FAPESP and CNPq, Brazil.
JF and FB would like to thank Ashok Das, D. G. C. McKeon  
and J. C. Taylor for many helpful discussions. 

\newpage 

\appendix 

\section{Three-loop graphs}
In this appendix we will present the results for the graphs in
figures \eqref{g3l1} and \eqref{g3l2} as well as some details of the calculation.
The first step of this calculation is rather straightforward involving
only a direct use of computer algebra
(all the calculations in this appendix as well as in the next,
have been performed using the computer algebra package HIP \cite{Hsieh:1991ti}). 
There are extra technical
difficults compared with similar calculations in QCD, which are
mainly associated with the trigonometric factors characteristic of NCQED.
As usual, in order to obtain the simplest possible expressions for the
three-loop graphs we have made use of kinematic identities in such a
way that the dot products $k_i \cdot k_j$ ($k_i$ represents any combination
of momenta) are reduced, whenever possible, to
quadratic terms such as  $k_i^2$ and $k_j^2$. 
The next step consists
in identifying which momentum shifts can be done so that one can
combine the momenta $k_i^2$ and $k_j^2$ as a single one, say
$k_i^2$. These shifts, however, are restricted to the ones which preserve
the trigonometric factors or reduce two or more trigonometric factors
to a single one.
%
%
This method is explained bellow in the case of the most complex graphs $(c)$ and
$(l)$, both containing four three-photon interaction vertices. Computation
for graphs $(h)$ is also detailed. 

We shall employ the following notation:
\begin{equation}
r =  p+k\, , \qquad s = q+k\, , \qquad t = p - q\, , \qquad 
\text{ s}(p,q) \equiv \sin\left(\frac{p\times    q}{2}\right) ,
\end{equation}
\begin{eqnarray}
\mathcal{T}_{1}  &\equiv & \text{ s}^2 (\,{k}, {p}\,) \,\text{ s}^2 (\,{k}, {q}\,)\nonumber \\  
\mathcal{T}_{2}  &\equiv & \text{ s} (\,{k}, {p}\,) \,\text{ s} (\,{k}, {q}\,)\,\text{ s} (\,{\it r}, 
{\it s}\,)\,\text{ s} (\,{p}, {q}\,)\nonumber \\  
\mathcal{T}_{3}  &\equiv & \text{ s} (\,{k}, {p}\,) \,\text{ s} (\,{k}, {q}\,)\,\text{ s} (\,p, 
 s\,)\,\text{ s}(\,{q}, {r}\,) . 
\end{eqnarray}
Before presenting the details, let us list the results.  
Denoting the contributions of each graph in figures \eqref{g3l1} 
and \eqref{g3l2} by $A_a \cdots A_l$, we obtain
%
%
\begin{subequations}
\bea
A_a &=& -12(d-1)^2 e^4\mathcal{T}_1 \left[\frac{1}{p^2 q^2 k^2 r^2}  + 2\frac{1}{k^4
    p^2 q^2} \right]\label{Ampa}\\ 
A_b &=& 4 e^4\mathcal{T}_1 \frac{d(d-1)^2}{k^4 p^2 q^2}\label{Ampb} \\
A_c &=& -e^4\mathcal{T}_1\bigg[\frac{4(6-5d)}{k^4 p^2 q^2} + \frac{(4d^2-38d+42)}{k^2p^2q^2r^2}-4(2d-3)^2\frac{(p\cdot q)^2}{k^4p^2 q^2r^2s^2} +
\frac{(6-11d-d^2)}{p^2q^2r^2s^2}\bigg]\label{Ampc} \\
A_d &=& 4(d-1)e^4\mathcal{T}_1\left[2\frac{1}{k^4 p^2 q^2} -
  \frac{1}{k^2 p^2 q^2 r^2} \right]\label{Ampd} \\
A_e &=& -2e^4\mathcal{T}_1  \left[ -2\frac{(p\cdot q)^2 }{k^4 p^2 q^2 r^2 s^2}+
  \frac{1 }{k^2 p^2 q^2 r^2} \right] \label{Ampe} \\
A_f &=& -e^4\mathcal{T}_1\bigg[8\frac{1}{k^4 p^2 q^2} -6(d-2)\frac{1}{k^2p^2q^2r^2}
-(d+2)\frac{1}{p^2q^2r^2s^2} + 8(2d-3)\frac{(p\cdot q)^2  }{k^4p^2
  q^2r^2s^2}  \bigg]\label{Ampf} \\
%
%
%
A_g &=& -2\frac{e^4\mathcal{T}_1}{p^2 q^2 r^2
    s^2} \label{Ampg} \\
A_h &=& -(d-1)e^4\left[18\frac{\mathcal{T}_1 +\mathcal{T}_2 }{p^2 q^2 r^2
    s^2} +4(2d-5)\mathcal{T}_2\frac{p\cdot (q+s)}{k^2p^2q^2r^2s^2}
\right]\label{Amph} \\
A_i &=& 2\frac{d(d-1)}{p^2 q^2 r^2 s^2}e^4(\mathcal{T}_1+\mathcal{T}_2)\label{Ampi} \\
A_j &=& -e^4\mathcal{T}_2\left[\frac{1}{p^2 q^2 r^2 s^2} -2\frac{p\cdot (q+s)  }{k^2
    p^2 q^2 r^2 s^2}\right]\label{Ampj} \\
A_k &=& -2 e^4\mathcal{T}_2\bigg[\frac{1}{p^2 q^2 r^2 s^2} + 2\frac{p\cdot
  (q+s)}{k^2p^2 q^2 r^2 s^2}\bigg]\label{Ampk} \\
A_l &=& -e^4\mathcal{T}_2\left[\frac{23-20d}{p^2q^2r^2s^2} -2(2d-5)
\frac{p\cdot(q+s)}{k^2p^2q^2r^2s^2}\right]. \label{Ampl} 
\eea
\end{subequations}

%
%

\subsection{The graph ${\bf c}$}

The amplitude for the graph $(c)$ has the form
\be
A_{c} \equiv -e^4\frac{\mathcal{T}_1}{k^4 p^2 q^2 r^2 s^2} F(\{(k_i\cdot
k_j)(k_i\cdot k_j)\})\, ,
\ee
where the function $F$ contains 148 terms belonging to the set
(This set itself has $155$ terms. $148$ is not the number of distinct
terms, rather, it is the number of unfactored pairs in terms of
dimension $d$, which may yield 2 or more polynomials in $d$ when no
numerical value is assigned to it. )
$$
\{(k_i\cdot k_j)(k_i\cdot k_j)\} = \{(p\cdot
k)(q\cdot r), (p\cdot q)k^2, \text{\dots}\}
$$
The first step to simplify $A_c$ consists in applying momentum
conservation only in non-quadratic terms, namely
$$
k\cdot r\, , k\cdot s\, , p\cdot r\, , p\cdot s\, , q\cdot r\, ,
q\cdot s\, , r\cdot s\, .
$$ 
Further, we employ the following identities: 
\be\label{krel1}
k\cdot p = \frac{1}{2}\left(r^2-p^2-k^2\right)\, , \qquad 
k\cdot q = \frac{1}{2}\left(s^2-q^2-k^2\right)\, . 
\ee
With this procedure the function $F$ can be expressed in terms of the much smaller set
$$
\{k_i^2 k_j^2\}; \qquad   (k_i , k_j ) \in \{r,s,p,q,k\} \, ,
$$
plus terms containing $p\cdot q$ and $(p\cdot q)^2$, which cannot be
simplified. There is  a total of $5!/(5-2)!2! +5 = 15$ pairs of the form
$k_i^2 k_j^2$ and certain ones are identical inside the integrand
after a change of variables. That is, they yield the same final
denominator and trigonometric factor. 

Next we analyze the equivalent pairs.
For example, a denominator of the  form
$$
\frac{1}{k^4 p^2 q^2}\, ,
$$
is generated by the binomials
$p^2 q^2$, $p^2 s^2$, $q^2 r^2$,  $r^2 s^2$.
That is, the first three terms can be made equivalent to the fourth
via the substitutions $p\lar -r, q \lar -s$, ${p\lar -r}$,
${q\lar -s}$, respectively.
For these pairs we obtain, respectively, the coefficients
$$
-\frac{3}{2}-d+2d^2 \, , \qquad -\frac{21}{2} + 11d-2d^2\,
,\qquad-\frac{21}{2} + 11d-2d^2\, ,\qquad -\frac{3}{2}-d+2d^2 \, 
$$
which combine to
\begin{equation}\label{d5}
4(6-5d)\frac{1}{k^4 p^2 q^2}\, .
\end{equation}
Next we have the equivalent terms
$k^2 r^2$, $k^2 s^2$, $k^2p^2$, $k^2 q^2$,
which are multiplied by the coefficients
$$
\frac{39}{2}+2d^2-17d\, , \qquad \frac{39}{2} + 2d^2-17d\, , \qquad
\frac{21}{2} - 2d^2-5d\, , \qquad \frac{21}{2} - 2d^2-5d\, . 
$$
Combining these terms, we obtain
\begin{equation}\label{ap4}
4(15-11d)\frac{1}{k^2p^2q^2r^2}.
\end{equation}
We also have $p^4$, $q^4$, $r^4$, $s^4$, all of them with
\begin{equation}\label{ap1}
-\frac{1}{2}(2d-3)(d-6)\, .
\end{equation}
And $p^2 r^2$, $q^2s^2$, with
\begin{equation}\label{ap2}
-(2d-3)(d-6)\, .
\end{equation}
Combining  \eqref{ap1} with \eqref{ap2} we obtain
\begin{equation}\label{ap3}
-\frac{1}{2}(2d-3)(d-6)\left[ \frac{(p^2-r^2)^2 +(q^2-s^2)^2}{k^4 p^2
    q^2 r^2 s^2}\right] \rightarrow -(2d-3)(d-6)\left[ \frac{(q^2-s^2)^2}{k^4 p^2 q^2 r^2 s^2}\right]
\end{equation}
Expanding the numerator of \eqref{ap3}
$$
(q^2-s^2)^2 = q^4 + s^4 -2q^2s^2 \rightarrow 2s^2(s^2-q^2) =
2s^2(k^2-k\cdot q) \rightarrow 2s^2 k^2\, ,
$$
we see that this term can be combined with that of \eqref{ap4},
yielding
\begin{equation}\label{ap9}
2\frac{12-7d-2d^2}{k^2p^2q^2r^2}
\end{equation}

The coefficient of the term $k^4$ of the function $F$ comes multiplied
by the coefficient 
\begin{equation}\label{ap7}
\frac{3}{2}-5d-d^2\, .
\end{equation}
Finally we have the term $p\cdot q$
\begin{eqnarray}\label{ap6}
2(2d-3)^2(p\cdot q)\frac{-2p\cdot q + p^2-r^2 +q^2-s^2
  +k^2}{k^4p^2q^2r^2s^2} &\rightarrow &  4(2d-3)^2(p\cdot q)\frac{-p\cdot q +q^2-s^2}{k^4p^2 q^2r^2s^2} \nonumber \\
  &&  + 2(2d-3)^2\frac{(p\cdot q)}{k^2p^2q^2r^2s^2} 
\end{eqnarray}
The term $p\cdot q (q^2 -s^2)$ can be simplified with the shift $p\lar -r$
\begin{eqnarray*}
\frac{p\cdot q (q^2 -s^2)}{k^4p^2 q^2r^2s^2} &\rightarrow& -\frac{p\cdot q (q^2 -s^2)}{k^4p^2 q^2r^2s^2} -\frac{q\cdot k (q^2 -s^2)}{k^4p^2 q^2r^2s^2}\\ \\
\frac{p\cdot q (q^2 -s^2)}{k^4p^2 q^2r^2s^2}&=&  -\frac{1}{4}\frac{(s^2-q^2-k^2) (q^2 -s^2)}{k^4p^2 q^2r^2s^2}
= -\frac{1}{4}\frac{\overbrace{(q^2 -s^2)}^{=0}}{k^2p^2 q^2r^2s^2} + \frac{1}{4}\frac{ (q^2 -s^2)^2}{k^4p^2 q^2r^2s^2} \\ \\
&=&\frac{1}{4}\frac{ q^4-2q^2s^2 +s^4}{k^4p^2 q^2r^2s^2}
\mapsto\frac{1}{2} \frac{ s^2(s^2-q^2)}{k^4p^2 q^2r^2s^2} =\frac{1}{2}\frac{(k^2+2k\cdot q)}{k^4p^2 q^2r^2} \rightarrow \frac{1}{2}\frac{1}{k^2p^2 q^2r^2} 
\end{eqnarray*}

In the last term of \eqref{ap6} we proceed as in \eqref{E10}, obtaining,
$ p\cdot q\rightarrow \frac{k^2}{4}$, and we finally have the
amplitude $A_c$
\begin{eqnarray}
A_c &=& -e^4\mathcal{T}_1\bigg[4(6-5d)\frac{1}{k^4 p^2 q^2} + (4d^2-38d+42)\frac{1}{k^2p^2q^2r^2}\nonumber \\ 
&&  \hspace*{0.5cm}-4(2d-3)^2\frac{(p\cdot q)^2}{k^4p^2 q^2r^2s^2} + (6-11d-d^2)\frac{1}{p^2q^2r^2s^2}\bigg]
\end{eqnarray}
The result for other graphs similar graphs $(e)$, $(f)$ and $(g)$ can
be obtained using the same procedure. The computation of graphs  $(a)$, $(b)$ and
$(d)$ is straightforward.

\subsection{The graph ${\bf h}$}

This graph as well as the graph $(i)$ are such that the output
expression contains combinations of the three trigonometric
factors. More specifically, $A_h = -e^4(h_{\mathcal{T}_1}\mathcal{T}_1 +
h_{\mathcal{T}_2}\mathcal{T}_2 +
h_{\mathcal{T}_3}\mathcal{T}_3)$, where
\begin{eqnarray*}
h_{\mathcal{T}_1} &=& 9k^2\\
h_{\mathcal{T}_2} &=& 2k^2 - 5k\cdot p - 5k\cdot q - 10 p\cdot q    + d[k^2  +
+ 2k\cdot p + 2 k\cdot q + 4p\cdot q ]\\
h_{\mathcal{T}_3} &=& k^2 + 5k\cdot p + 5k\cdot q + 10 p\cdot q    - d[k^2  +
+ 2k\cdot p + 2 k\cdot q + 4p\cdot q ]\\
\end{eqnarray*}
After the substitutions $k\cdot p = \frac{1}{2}[r^2-p^2-k^2]$,
$k\cdot q= \frac{1}{2}[r^2-p^2-k^2]$ we obtain
\begin{eqnarray}
A_h &=& -e^4\frac{d-1}{k^2p^2q^2r^2s^2}\bigg [2 (9 \mathcal{T}_1 +(d+2)\mathcal{T}_3
-(d-7)\mathcal{T}_2)k^2 + 4(2d-5) (\mathcal{T}_2-\mathcal{T}_3)p\cdot q\nnbb
&&\hspace*{2.3cm}+ (2d-5)(\mathcal{T}_2-\mathcal{T}_3)(r^2+s^2-p^2-q^2)\bigg]
\end{eqnarray} 
Collecting terms proportional to $k^2$
\begin{equation}\label{B1}
h_{k^2} \equiv  2\frac{(d-1)}{p^2 q^2 r^2 s^2}\left[9\mathcal{T}_1
  -(d-7)\mathcal{T}_2 
+(d+2)\mathcal{T}_3 \right] .
\end{equation}
The $d$-independent part in brackets can be eliminated via ${\mathcal T}_{3} =
(p\leftrightarrow -r,{\mathcal T}_{2})$, which allows us to write
\begin{eqnarray}
h_{k^2} &=&  2\frac{(d-1)}{p^2 q^2 r^2 s^2}\left[9\mathcal{T}_1 +(d+2)\frac{\mathcal{T}_3+\mathcal{T}_2}{2}-(d-7)\frac{\mathcal{T}_2+\mathcal{T}_3}{2} \right] \nonumber \\ \nonumber \\
&=& 9\frac{(d-1)}{p^2 q^2 r^2 s^2} \left[2\mathcal{T}_1 +\mathcal{T}_2
  + \mathcal{T}_3 \right]\rightarrow 18\frac{(d-1)}{p^2 q^2 r^2 s^2}
\left[\mathcal{T}_1 + \mathcal{T}_2\right]
\end{eqnarray}
The terms proportional to $p\cdot q$ give
\begin{eqnarray}\label{B2}
h_{p\cdot q} &\equiv& 4\frac{(d-1)(2d-5)}{k^2p^2q^2r^2s^2} (\mathcal{T}_2-\mathcal{T}_3)p\cdot q\nonumber \\ \nonumber \\
&\rightarrow& 4\frac{(d-1)(2d-5)}{k^2p^2q^2r^2s^2} \mathcal{T}_2 p\cdot (q+s)
\end{eqnarray}
Collecting the remaining terms, we have
\begin{eqnarray}\label{B3}
h_{rem} &\equiv& (\mathcal{T}_2-\mathcal{T}_3)\frac{(d-1)(2d-5)}{k^2p^2q^2r^2s^2} (r^2+s^2-p^2-q^2) \nonumber \\ \nonumber \\
&=& -2(\mathcal{T}_2 - \mathcal{T}_3)\frac{(d-1)(2d-5)}{k^2p^2q^2r^2s^2} (p^2+q^2)\nonumber \\ \nonumber \\
&\rightarrow& -4(\mathcal{T}_2 - \overbrace{\mathcal{T}_3}^{p\lar- r})\frac{(d-1)(2d-5)}{k^2p^2r^2s^2} \rightarrow 0 
\end{eqnarray}
\begin{eqnarray}\label{B4}
A_h &=& -(d-1)e^4\left[18\frac{\mathcal{T}_1 +\mathcal{T}_2 }{p^2 q^2 r^2 s^2} +4(2d-5)\mathcal{T}_2\frac{p\cdot (q+s)}{k^2p^2q^2r^2s^2} \right]
\end{eqnarray}

\subsection{The graph ${\bf l}$}

After proceeding as we did with the graph $(c)$, we obtain
\begin{eqnarray}
A_l &=& -e^4\frac{\mathcal{T}_2}{k^2 p^2 q^2 r^2 s^2 t^2} \bigg\{
-6(d-1)(q^2 r^2 + t^2 k^2 + s^2 p^2 )
+\frac{(2d-5)}{3}\bigg[k^4+p^4+q^4+s^4+t^4 \nnbb
&& -q^2k^2-q^2p^2-q^2s^2-q^2t^2-t^2p^2-t^2r^2-t^2s^2-r^2s^2-r^2p^2-r^2k^2-s^2k^2-p^2k^2) \bigg]\bigg\}
\end{eqnarray}
This result can be simplified by the shifts $(p\leftrightarrow q,r\leftrightarrow s)$ to
\begin{eqnarray}\label{E1}
A_l &=& 
 -e^4\mathcal{T}_2\bigg\{-6(d-1)\left[ \frac{2}{k^2p^2s^2t^2} + \frac{1}{p^2q^2r^2s^2}\right]\nonumber   \\ \nonumber \\
&&\hspace*{0.8cm} +\frac{(2d-5)}{3} \left[ \frac{k^2}{p^2q^2r^2s^2t^2} + \frac{2p^2}{k^2q^2r^2s^2t^2} + \frac{2r^2}{k^2p^2q^2s^2t^2} + \frac{t^2}{k^2p^2q^2r^2s^2}\right] \nonumber \\ \nonumber \\
 &&\hspace*{0.8cm} -\frac{(2d-5)}{3} \left[ \frac{2}{p^2 r^2 s^2 t^2}  + \frac{2}{k^2 p^2 r^2 t^2} + \frac{2}{k^2 p^2 r^2 s^2 } +  \frac{2}{k^2 p^2 q^2 s^2 } + \frac{2}{p^2 q^2 s^2 t^2} +\frac{1}{k^2 p^2 q^2 t^2} +   \frac{1}{k^2 r^2 s^2 t^2}\right] \bigg\}\nonumber \\ \nonumber \\
&\equiv& -e^4(L_1+L_2+L_3)\, ,
\end{eqnarray}
where the labels $L_i$, correspond to the each line of \eqref{E1}. 

The results for  $L_1$ and $L_2$ and $L_3$ are
\begin{eqnarray}\label{E8}
L_1 &=& -18\mathcal{T}_2\frac{(d-1)}{p^2q^2r^2s^2} \nonumber \\ \nonumber \\
L_2&=& 4(2d-5) \mathcal{T}_2 \bigg[  \frac{1}{k^2p^2r^2s^2} -\frac{p\cdot q}{k^2p^2q^2r^2s^2} \bigg] \nonumber \\ \nonumber \\
L_3 &=&  -\mathcal{T}_2\frac{(2d-5)}{3} \bigg[ \frac{2}{p^2 r^2 s^2 t^2}  + \frac{2}{k^2 p^2 r^2 t^2}  +  \frac{2}{k^2 p^2 q^2 s^2 }+\frac{2}{k^2 p^2 r^2 s^2 } + \frac{2}{p^2 q^2 s^2 t^2} + \overbrace{\frac{1}{k^2 p^2 q^2 t^2}}^{(p\leftrightarrow -r),(q\leftrightarrow -s)} +   \frac{1}{k^2 r^2 s^2 t^2}\bigg]\nonumber \\ \nonumber \\
&=&-2\mathcal{T}_2\frac{(2d-5)}{3} \bigg[ \frac{1}{p^2 r^2 s^2 t^2}  + \frac{1}{k^2 p^2 r^2 t^2}  +  \frac{1}{k^2 p^2 q^2 s^2 }+\overbrace{\frac{1}{k^2 p^2 r^2 s^2 }}^{k\leftrightarrow -p} + \frac{1}{p^2 q^2 s^2 t^2}  +   \overbrace{\frac{1}{k^2 r^2 s^2 t^2}}^{k\leftrightarrow -p}\bigg]\nonumber \\ \nonumber \\
&=&-2\mathcal{T}_2\frac{(2d-5)}{3} \bigg[ \frac{2}{p^2 r^2 s^2 t^2}  + \frac{2}{k^2 p^2 r^2 t^2}  +  \overbrace{\frac{1}{k^2 p^2 q^2 s^2 }}^{(p\leftrightarrow t),(k\leftrightarrow -s)} + \frac{1}{p^2 q^2 s^2 t^2} \bigg]\nonumber \\ \nonumber \\
&=&-2\mathcal{T}_2\frac{(2d-5)}{3} \bigg[ \frac{2}{p^2 r^2 s^2 t^2}  + \frac{2}{k^2 p^2 r^2 t^2}  +  \overbrace{\frac{1}{k^2 q^2 s^2 t^2 } + \frac{1}{p^2 q^2 s^2 t^2}}^{p\leftrightarrow q} \bigg]\nonumber \\ \nonumber \\
&=&-2\mathcal{T}_2\frac{(2d-5)}{3} \bigg[ \frac{2}{p^2 r^2 s^2 t^2}  +
\frac{3}{k^2 p^2 r^2 t^2}  +\overbrace{  \frac{1}{p^2 q^2 r^2
    t^2}}^{(p\leftrightarrow -r),(q\leftrightarrow -s)} \bigg] = -2\mathcal{T}_2(2d-5) \bigg[ \frac{1}{p^2 r^2 s^2 t^2}  + \overbrace{  \frac{1}{k^2 p^2 r^2 t^2} }^{(p\leftrightarrow t),(k\leftrightarrow -s)} \bigg]\nonumber \\ \nonumber \\
&=&-4 \overbrace{\frac{(2d-5)}{p^2 r^2 s^2 t^2}\mathcal{T}_2}^{(q=-s,k=-r)} = -4 \overbrace{\frac{(2d-5)}{k^2 q^2 r^2 p^2}\mathcal{T}_2}^{q=-s} = -4 \frac{(2d-5)}{k^2 p^2 r^2 s^2}\mathcal{T}_3
\end{eqnarray}

Finally
\begin{eqnarray}\label{E9}
A_l &=& -e^4(L_1 + L_2 + L_3) \nonumber \\ \nonumber \\
    &=& e^4\left[
18\frac{(d-1)}{p^2q^2r^2s^2}\mathcal{T}_2 + 4\frac{(2d-5)}{k^2 p^2 r^2 s^2}(\mathcal{T}_2-\mathcal{T}_3) -4(2d-5)\frac{p\cdot q}{k^2p^2q^2r^2s^2}\mathcal{T}_2\right] \nonumber \\ \nonumber \\
    &=& e^4\left[
18\frac{(d-1)}{p^2q^2r^2s^2}\mathcal{T}_2  -4(2d-5)\frac{p\cdot q}{k^2p^2q^2r^2s^2}\mathcal{T}_2\right] ,
\end{eqnarray}
where we have used
$$
\frac{(\mathcal{T}_2-\overbrace{\mathcal{T}_3}^{p\leftrightarrow -r})}{k^2 p^2 r^2 s^2} = 0\,.
$$

Notice that, in the absence of the trigonometric factor (i.e., as in
QCD) the second term of \eqref{E9} can be simplified by the relation
\begin{eqnarray}\label{E10}
\int \frac{\overbrace{p}^{p\rightarrow -r}\cdot q}{k^2p^2q^2r^2s^2} &=& -\int \frac{p \cdot q}{k^2p^2q^2r^2s^2}  -\int \frac{k \cdot q}{k^2p^2q^2r^2s^2}\nonumber\\ \nonumber \\
\int \frac{p\cdot q}{k^2p^2q^2r^2s^2} &=& -\frac{1}{2}\int
\frac{\overbrace{k \cdot q}^{k \cdot q=\frac{1}{2}s^2 -\frac{1}{2}q^2
    - \frac{1}{2}k^2}}{k^2p^2q^2r^2s^2} =  \frac{1}{4}\int \frac{1}{p^2 q^2 r^2 s^2}
\end{eqnarray}
Using the same procedure in the noncommutative case one finds 
a more complicated expression, however, this step is necessary to the
final result.

Shifting $p\rightarrow -r$ and $q\rightarrow -s$, $\mathcal{T}_2\rightarrow \mathcal{T}_3$
we write the symmetrization
\begin{eqnarray}\label{E11}
\mathcal{T}_2 \frac{p\cdot q}{k^2 p^2 r^2 s^2} &=& \frac{1}{4}\bigg[ \frac{\mathcal{T}_2 p\cdot q}{k^2 p^2 r^2 s^2} + \frac{\overbrace{\mathcal{T}_2 p}^{p\leftrightarrow -r}\cdot q}{k^2 p^2 r^2 s^2}  + \frac{p\cdot \overbrace{\mathcal{T}_2 q}^{q\leftrightarrow -s}}{k^2 p^2 r^2 s^2} + \frac{\overbrace{\mathcal{T}_2 p\cdot q}^{p\leftrightarrow -r,q\leftrightarrow -s}}{k^2 p^2 r^2 s^2}\bigg] \nonumber \\ \nonumber \\
&=& \frac{1}{4}\bigg[ \mathcal{T}_2\frac{ p\cdot q}{k^2 p^2 r^2 s^2} -\mathcal{T}_3\frac{ r\cdot q}{k^2 p^2 r^2 s^2}  - \mathcal{T}_3\frac{p\cdot s}{k^2 p^2 r^2 s^2} + \mathcal{T}_2\frac{ r\cdot s}{k^2 p^2 r^2 s^2}\bigg] 
\end{eqnarray}
Then we finally substitute $r=p+k$, $s=q+k$ and obtain
\begin{eqnarray}\label{E12}
\hspace*{-3.2cm}\mathcal{T}_2 \frac{p\cdot q}{k^2 p^2 r^2 s^2} &=& \frac{1}{2}\left[\frac{1}{2}\frac{1}{p^2q^2r^2s^2} + \frac{p\cdot(q+s)}{k^2p^2q^2r^2s^2}\right]
\end{eqnarray}
\begin{eqnarray}
A_l &=& -e^4\left[\frac{23-20d}{p^2q^2r^2s^2} -2(2d-5)\frac{p\cdot(q+s)}{k^2p^2q^2r^2s^2}\right]\mathcal{T}_2
\end{eqnarray}
The other {\it Mercedes} graphs, $(j)$ and $(k)$ can be dealt with in
the same fashion.

\section{Gauge parameter (in-) dependence}\label{appB}

\subsection{Two-loops}
From the structure of the graphs in figure \eqref{1fig2} we expect a
polynomial of third degree in $\bar\xi$. The highest power comes only from
the graph containing three photon propagators. A direct calculation
shows that such a contribution is identically zero. This can be
understood as a consequence of the Ward identity for the cubic vertex
\be\label{1ward1}
k^\mu\,\left(
\begin{array}{c}\includegraphics[scale=0.8]{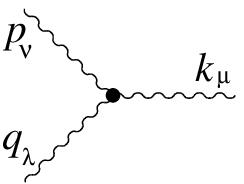}\end{array} \right)
= -2\,i\,e\,\sin\left(\frac{p \times q}{2}\right) \left[
 \left(p_\nu\,p_\lambda  -  p^2 \eta_{\nu\lambda}\right) 
-\left(q_\nu\,q_\lambda  -  q^2 \eta_{\nu\lambda}\right) 
\right]
\ee
and the structure of the gauge dependent part of the photon propagator in Eq. \eqref{1f1}.
Then the vanishing of the $\bar\xi^3$ terms follows from the
transversality of \eqref{1ward1}.

The quadratic terms involve contributions from
the first and the second graphs in figure \eqref{1fig2}.
Performing shifts and using relations like \eqref{krel1} the individual contributions of
these two graphs can be reduced to
\begin{eqnarray*}
G_1^{(\bar{\xi}^2)} &=& e^2 s^2(p,q) \left(\frac{(p\cdot q)^2}{p^4 q^4} -
  \frac{1}{p^2 q^2} \right)\\
G_2^{(\bar{\xi}^2)} &=& -e^2 s^2(p,q) \left(\frac{(p\cdot q)^2}{p^4 q^4} -
  \frac{1}{p^2 q^2}\right)
\end{eqnarray*}
which clearly cancels each other. This simple cancellation only occurs
because we have the freedom to perform shifts inside the 
regularized integrals.

The contributions which are linear in $\bar\xi$ are generated by all
the three graphs in figure \eqref{1fig2}. Proceeding similarly as in
the case of the quadratic contributions, the individual graphs can be
reduced to the following expressions
\begin{eqnarray*}
G_1^{(\bar{\xi})} &=& -e^2\left[
\frac{3}{2}\frac{s^2(p,q)}{p^2 q^2} + 2d\frac{s^2(p,q)}{p^2 q^2}\right]\, \\ 
G_2^{(\bar{\xi})} &=& e^2\left[2\frac{s^2(p,q)}{p^2 q^2} - 
2d\frac{s^2(p,q)}{p^2 q^2}\right]\, \\ 
G_3^{(\bar{\xi})} &=& -\frac{e^2}{2}\frac{s^2(p,q)}{p^2 q^2}
\end{eqnarray*}
so that $G_1^{(\bar{\xi})}+G_2^{(\bar{\xi})}+G_3^{(\bar{\xi})} = 0$.
This concludes the verification of the gauge parameter independence of
the free energy at two-loop order.

\subsection{Three-loops}

Unlike the two-loop contributions,
in the present case one does not expect a gauge parameter independence.
However, there must be some cancellation, so that the residual
dependence will combine with the perturbative corrections to the coupling
constant, which are not included in the present analysis.

Indeed, the straightforward computer algebra calculation shows that the $\bar{\xi}^6$ 
contribution (from the graphs $(c)$ and $(l)$) 
as well as $\bar{\xi}^5$ (from graphs $(a)$, $(c)$,
$(h)$ and $(l)$) vanish. This is also a direct consequence of the Ward
identities like \eqref{1ward1} (as well as the analogous identity
involving the three and four photon vertices) and holds at the integrand level.
We hale also verified that at order $\bar{\xi}^4$ every graph with 
four or more photon propagators gives a non-zero contribution, except $(f)$. 

We have taken advantage of the fact we are working in a space-time of $d$ 
dimensions in order to organize the gauge dependent contributions
according to the power of $d$. Our calculation shows that the 
highest power of $d$ is three, so that an arbitrary  amplitude can be written as
\be
A^{\rm 3-loops} = e^4\left(A_0 + d A_1 + d² A_2 + d³ A_3\right)\, .
\ee
Since $d$ can be arbitrary we can study gauge parameter dependence for each
power of $d$ as well as $\bar{\xi}$. Usually the highest power of the
dimension will have less terms to deal with, so that in order 
to understand how these terms combine we will study two simple cases,
namely order $\bar{\xi}$ and $\bar{\xi}^2$ in the gauge parameter and $d^2$ in the dimension.

\subsubsection{Order $(\bar{\xi}^2,d^2)$}

Only the graphs $(a)$, $(b)$ and $(h)$ have non-zero contributions at
$(\bar{\xi}^2, d^2)$. They all share the same trigonometric factor,
$\mathcal{T}_1$, which is invariant under $q\leftrightarrow p$ and
shifts $q\rightarrow\pm q-k$, $p\rightarrow\pm p-k$. Keeping in mind
we'll just use these transformations, we will 
omit $e^4\mathcal{T}_1$, so that the resulting amplitudes are

\begin{eqnarray}
A_a^{(\bar{\xi}^2,d^2)} &=& 4\frac{(k^2 +2 p\cdot k)^2}{(k^2)^3 q^2 p^2 (p+k)^2} \\ 
A_b^{(\bar{\xi}^2,d^2)} &=& -4 \frac{1}{(k^2)^2 q^2 p^2} \\ 
A_h^{(\bar{\xi}^2,d^2)} &=& -16 \frac{(k^2 +2 q\cdot k)^2 (k^2 +2 p\cdot k)^2 }{(k^2)^4 q^2 p^2 (k+p)^2 (k+q)^2}
\end{eqnarray}

After expanding each amplitude and using \eqref{krel1} we obtain

\begin{eqnarray}
A_a^{(\bar{\xi}^2,d^2)} &=& 8\frac{(p+k)^2}{(p^2)^3 q^2 k^2}-8\frac{1}{(p^2)^3 q^2} \\ 
A_b^{(\bar{\xi}^2,d^2)} &=& -4 \frac{1}{(p^2)^2 q^2 k^2} \\
A_h^{(\bar{\xi}^2,d^2)} &=& 8 \frac{(p+q)^2}{(p^2)^4 q^2} 
-4\frac{(p+q)^2 (p+k)^2}{(p^2)^4 q^2 k^2} -4\frac{1}{(p^2)^4}
\end{eqnarray}
Discarding the terms which are odd functions of the momenta
(keeping in mind that $\mathcal{T}_1$ is an even function) we have
\be
A_a^{(\bar{\xi}^2,d^2)}  \rightarrow 8 \frac{1}{(p^2)^2 q^2 k^2}
\ee

Analogously, we take the even part of $A_d$
\be
A_d^{(\bar{\xi}^2,d^2)} 
\rightarrow \frac{1}{4}\left[A_d(p,q,k) + A_d(p,q,-k) + A_d(p,-q,k)  +
  A_d(p,-q,-k) \right]  
=-4 \frac{1}{(p^2)^2 q^2 k^2}\, .
\ee
Finally
\be
A_a^{(\bar{\xi}^2,d^2)}  + A_c^{(\bar{\xi}^2,d^2)}  + A_d^{(\bar{\xi}^2,d^2)}  =0\, .
\ee

\subsubsection{Order $(\bar{\xi},d^2)$}

The graphs $a$, $b$, $c$ and $h$ contribute to order $(\bar{\xi}, d^2)$. In this case we have to deal with a more complicated trigonometric factor for the graph $(h)$, which has the amplitude

\begin{eqnarray}
A_h^{(\bar{\xi},d^2)} &=& 2 e^4
\frac{s(k,p) s(k,q) \left[ s(q,p) s(r,s) + s(q,r)s(p,s)\right]}{p^2 q^2 k^2 r^2 s^2} (r\cdot k + q\cdot k)(s\cdot k + p\cdot k)\nnbb
&\times& \delta(s-k-q)\delta(r-k-p)
\end{eqnarray}

We shall only be concerned with transformations which preserve the delta functions, namely
\begin{equation}\label{inv21}
r\lar -p \, , \qquad s\lar -q\, , \qquad \{s\lar r\, ,q\lar p\}\, .
\end{equation}
Keeping in mind that we will be only using these transformations we will always
write  $s(r,p)=s(k,p)$, etc.
Shifting $r\lar -p$ in the first term in brackets and his multiples, we have
\be
A_h^{(\bar{\xi},d^2)} = 4 \frac{s(k,p) s(k,q) s(q,r) s(p,s)}{p^2 q^2 k^2 r^2 s^2} (r\cdot k + q\cdot k)(s\cdot k + p\cdot k)
\ee
This expression has a trigonometric factor which is invariant 
by the transformations \eqref{inv21}. Omiting $e^4 {\cal T_3}$
and using the delta functions we obtain
\begin{eqnarray}
A_h^{(\bar{\xi},d^2)} &\rightarrow& -8 \frac{k\cdot q}{p^2 q^2 
k^2 (p+k)^2(q+k)^2} -8 \frac{p\cdot k}{p^2 q^2 k^2 (p+k)^2 (q+k)^2} \nnbb
&&  -16 \frac{(k\cdot q)( k\cdot p)}{(p^2)^2 q^2 k^2 (p+q)^2(p+k)^2} - 4\frac{1}{ q^2 k^2 (p+q)^2 (p+k)^2}
\nnbb
&\rightarrow &  -8 \frac{1}{k^4 q^2 p^2 } +8 \frac{1}{k^4 q^2 (p+k)^2 }
\end{eqnarray}
where the last line is obtained with identities such as \eqref{krel1}.

The amplitude from graph $(c)$ has 20 terms. After shifts they are reduced to
\begin{eqnarray}
A_d^{(\bar{\xi},d^2)}  &\rightarrow& -8 \frac{p\cdot q}{(k^2)^3 q^2
  (p+k)^2} +8 \frac{p\cdot q}{(k^2)^3 (p+k)^2 (q+k)^2} \nnbb
&&  +8 \frac{p\cdot q}{(k^2)^3 q^2 p^2} - 8 \frac{p\cdot q}{(k^2)^3 (k+q)^2 p^2}\rightarrow 8 \frac{1}{k^4 q^2 k^2}
\end{eqnarray}
Likewise we have for graphs $(a)$ and $(b)$
\begin{eqnarray}
A_a^{(\bar{\xi},d^2)} &\rightarrow& -8 \frac{1}{((p+k)^2)^2 q^2 k^2 } +8 \frac{1}{k^2 q^2 p^4 } - 24\frac{1}{k^4 q^2 p^2 } \\ 
A_c^{(\bar{\xi},d^2)} &\rightarrow&  16\frac{1}{k^4 p^2 q^2} 
\end{eqnarray}
Combining every individual result we have
\begin{eqnarray}
A^{(\bar{\xi},d^2)} &=& \mathcal{T}_1\left[8 \frac{1}{k^2 q^2 p^4} - 8 \frac{1}{((p+k)^2)^2 q^2 p^2 } \right] + \mathcal{T}_2 \left[8 \frac{1}{k^4 q^2 p^2 } - 8 \frac{1}{k^4 q^2 (p+k)^2 }\right]\, ,   
\end{eqnarray}
Next we make shifts to write each denominator as one, 
$\frac{\dst 1}{\dst k^4 p^2 q^2}$\, , and sum the respective trigonometric factors, which have been altered
\begin{eqnarray}
S_1 &\equiv& s(p,q)² s(p,k)²\\ 
S_2 &\equiv& -s(k-p,q)² s(k,p)²\\ 
S_3 &\equiv& s(k,p) s(k,q) s(q,p+k) s(p,q+k) \\
S_4 &\equiv& s(k,p) s(k,q) s(p,q) s(p-k,k+q)
\end{eqnarray}
Taking the even part of $S_T \equiv S_1+S_2+S_3+S_4$ in the variable $q$, we have 
\be
\frac{S_T(p,q,k)+S_T(p,-q,k)}{2} = -2 s(k,q) c(q,p) c(q,k) s(q,p) + 2 s(k,q) c(q,p) c(q,k) s(q,p) c(k,p)²\, ,
\ee
which is odd in $k$ and $p$, so  
\be
\int d^d k \frac{S_T}{k^4 p^2 q^2} = \frac{1}{2}\int d^d k \frac{S_T(p,q,k)+S_T(p,-q,k)}{k^4 p^2 q^2} = 0.
\ee

This calculation shows how the gauge parameter dependence of the
three-loop graphs can be reduced to a smaller class of terms.
The remaining gauge parameter dependence must cancel when we
take into account the corrections to the coupling constant. 
However, we have verified that 
all these contributions are sub-leading in the high temperature
regime, so that only the gauge invariant ring contributions survive.

\end{document}